\documentclass[12pt]{article}
\usepackage{graphicx}
\input epsfig.sty
\newsavebox{\sboxpubnumber}
\newsavebox{\sboxpubdate}

\newcommand{\pubnumber}[1]{\begin{lrbox}{\sboxpubnumber}{\begin{tabular}{l} #1 \\
				 \usebox{\sboxpubdate}
				 \end{tabular}}
                           \end{lrbox}
                           \pubblock}
\newcommand{\Title}[1]{\begin{center} {\Large #1 } \end{center}}
\newcommand{\Author}[1]{\begin{center}{ \sc #1} \end{center}}
\newcommand{\Address}[1]{\begin{center}{ \it #1} \end{center}}

\newcommand{\pubblock}{\rightline{
			\usebox{\sboxpubnumber}}}
\newenvironment{Abstract}{\begin{quotation}  }{\end{quotation}}

\newcommand{\Acknowledgements}{\bigskip  \bigskip \begin{center} 
\begin{large}
             \bf ACKNOWLEDGEMENTS \end{large}\end{center}}
\def\laq{\raise 0.4ex\hbox{$<$}\kern -0.8em\lower 0.62 ex\hbox{$\sim$}}
\def\gaq{\raise 0.4ex\hbox{$>$}\kern -0.7em\lower 0.62 ex\hbox{$\sim$}}

\begin{document}
\begin{titlepage}
\pubnumber{UNIL-IPT 02-09} 
\vfill
{}\Title{PRIMORDIAL MAGNETIC FIELDS}
\vfill
\Author{Massimo Giovannini\footnote{Electronic address: 
Massimo.Giovannini@ipt.unil.ch, Massimo.Giovannini@cern.ch}}
\Address{Institute of Theoretical Physics, University of Lausanne, \\
BSP CH-1015, Dorigny, Switzerland\\
and\\
Theoretical Physics Division, CERN, CH-1211, Geneva, Switzerland}
\vfill
\begin{Abstract}
Large scale magnetic fields represent a
triple point where cosmology, high-energy physics and 
astrophysics meet for different but related purposes. 
After reviewing the implications of large scale 
magnetic fields in these different areas, the 
r\^ole of primordial magnetic fields is discussed 
in various physical processes occurring prior to the 
decoupling epoch with particular attention to the 
big bang nucleosynthesis (BBN) epoch and to the 
electroweak (EW) epoch. The generation  
of matter--antimatter isocurvature fluctuations, 
induced by hypermagnetic fields, is analyzed
in light of a possible increase of extra-relativistic species
at BBN. It is argued that stochastic GW backgrounds 
can be generated by hypermagnetic fields at the LISA 
frequency. The problem 
of the origin of large scale magnetic fields is also
scrutinized.
\end{Abstract}
\vfill
\centerline{\sl Prepared for}
\centerline{\sl 7th ``Colloque de Cosmologie''}
\centerline{\sl     ``High-Energy Astrophysics for and from space'}
\centerline{   Paris,  June 11 -- 15, 2002.}
\centerline{To appear in the Proceedings}
\vfill
\end{titlepage}
\def\thefootnote{\fnsymbol{footnote}}
\setcounter{footnote}{0}
%%%%%%%%%%%%%%%%%%
\renewcommand{\theequation}{1.\arabic{equation}}
\setcounter{equation}{0} 
\section{A triple point}
Through 
the last fifty years, the possible existence and implications  of primordial 
magnetic fields became a very useful 
cross-disciplinary area at the interface of cosmology, 
astrophysics and high-energy physics\cite{rev1,rev2,rev3,rev4,rev5,rev6,rev7}. 
From astrophysical observations, we do know that planets, 
stars, the interstellar medium and the intergalactic medium
 are all magnetized. The 
magnetic fields in these environments have values ranging from the 
$\mu $ G of the intra-cluster medium, to 
the G  (in the case of the earth) up to presumably $10^{12}$ G, the 
typical magnetic fields of neutron stars. In principle, 
observations of magnetic fields in galaxies (and clusters 
of galaxies) could discriminate between a direct
primordial origin of the large-scale fields and a primordial 
origin mediated by a dynamo amplification. None of the 
two options are, at the moment, supported by clear 
observational evidence. Furthermore, in both 
approaches there  are various theoretical assumptions 
which have been (but still need to be) carefully 
scrutinized.

In high-energy physics, the possible existence 
of intergalactic magnetic fields is one of the crucial 
unknowns in the analysis of ultra-high energy cosmic rays 
above the GZK cut-off. The interplay of high-energy physics 
and astrophysics is indeed present since the origin 
of this subject. In 1949 the  scientific argument
between Fermi \cite{fermi}, on one side,and Alfv\'en \cite{alv1,alv2},
 Richtmyer and  Teller \cite{alv3}, on the other, 
concerned exactly the possible existence of 
galactic magnetic fields. Fermi was convinced that high-energy 
cosmic rays are in equilibrium with the whole 
galaxy while Alfv\'en was supporting the 
idea that high energy cosmic rays are in equilibrium with stars. 
In order to make his argument consistent, Fermi postulated (rather than 
demonstrated) the existence of a $\mu$ G galactic magnetic 
field. Fermi thought that the origin of this field was  primordial.

In cosmology the possible 
existence of magnetic fields prior to decoupling can 
influence virtually all the moments in the thermodynamical history 
of the Universe. Big-bang nucleosynthesis (BBN), 
electroweak phase transition (EWPT), decoupling time are all influenced 
by the existence of magnetic fields at the corresponding epochs. 
If magnetic fields were originated in the 
past history of the Universe, their birth should be related, in some 
way, to the interplay of gravitational and gauge interactions. 
Superstring theories and higher-dimensional theories, formulated 
 through the past thirty years, pretend 
to give us some hints on the possible form of such an interplay and 
their implications may be useful to consider. 

The present paper is organized as follows. 
In Section II the basic ideas on large scale 
magnetic field structure and observations will 
be briefly outlined. 
Section III collects some considerations 
on the evolution of magnetic fields. In Section 
IV the problem of the origin will be 
illustrated with particular attention to models 
where there is an effective evolution of the 
gauge coupling. Section V deals with the 
possible implications of hypermagnetic fields 
for the EW physics and for the generation of the BAU.
In Section VI it will be shown that if hypermagnetic fields 
are present at the EW epoch, matter--antimatter
fluctuations are likely to be produced at BBN. 
In Section VII the implications of hypermagnetic 
fields for the GW backgrounds at the LISA 
and VIRGO/LIGO frequencies will be discussed. 
In Section VIII some speculations 
on the possible Faraday rotation of the CMB polarization 
will be presented. Section IX contains some concluding remarks.

\renewcommand{\theequation}{2.\arabic{equation}}
\setcounter{equation}{0} 
\section{A primer on magnetic fields observations}

There already excellent reviews on the measurements 
of large scale magnetic fields in diffuse  astrophysical plasmas 
\cite{rev1,rev2,rev3,rev4,rev5}. Here we want just to give 
an elementary primer on  the 
main ingredients of the detection strategies 
and focus the attention on recent observations.

\subsection{ Local and global observables} 

In order to measure large scale magnetic fields, 
one of the first effects coming to mind,  is  the Zeeman splitting.
The energy levels of  an hydrogen atom in the background of a magnetic field 
are not degenerate. The presence of a magnetic field
produces a well known splitting of the spectral lines:
\begin{equation}
\Delta \nu_{Z} = \frac{e B_{\parallel}}{2 \pi m_{e}}.
\label{zee}
\end{equation}
From the estimate of the 
splitting, the magnetic field intensity can be deduced. 
Now, the most common element in the interstellar medium 
is neutral hydrogen, emitting the famous $21$-cm 
line (corresponding to a frequency of $1420$ MHz). 
Suppose that a magnetic field of $\mu$ G strength is present 
in the interstellar medium. From 
Eq. (\ref{zee}), the induced splitting,
 $\Delta\nu_{Z} \sim 3 {\rm Hz}$, can be estimated. 
Hence, Zeeman splitting of 
the $21$-cm line generates two oppositely circular 
polarized spectral lines whose apparent splitting is 
however sub-leading if compared to the Doppler broadening.
In fact, the atoms and molecules in the interstellar medium 
suffer thermal motion and the 
amount of induced Doppler broadening is roughly given by 
\begin{equation}
\Delta \nu_{\rm Dop} \sim \biggl(\frac{v_{\rm th}}{c}\biggr) \nu,
\label{dop}
\end{equation}
where $ v_{\rm th}$ is the thermal velocity $\propto \sqrt{T/m}$ where 
$m$ is the mass of the atom or molecule. The amount of Doppler broadening 
is $\Delta \nu_{\rm Dop} \sim 30$ kHz which is much larger than the 
Zeeman splitting we ought to detect.
Zeeman splitting of molecules and recombination lines should however 
be detectable if the magnetic field strength gets larger with the density. 
Indeed in the interstellar medium there are molecules with an unpaired 
electron spin. In these cases a Zeeman 
splitting comparable with the one of the $21$-cm line can be 
foreseen. These 
molecules include OH, CN, CH and some other. In the past, OH clouds 
were used in order to estimate the magnetic field (see \cite{rev3} and references therein). 
The possible 
caveat with this type of estimates is that the measurements can 
only be very {\em local}. The above mentioned molecules are much less common 
than neutral hydrogen and are localized in specific 
regions of the interstellar medium. 

The first experimental evidence of the existence of large scale magnetic 
fields in galaxies came from the synchrotron emission.  The emissivity formula 
for the synchrotron depends upon $B_{\perp}$ and upon the relativistic 
electron density. The synchrotron has an intrinsic polarization 
which can give the orientation of the magnetic field, 
 but not the specific sign of 
the orientation vector. The relativistic electron density is sometimes 
estimated using equipartition, i.e. the idea that 
magnetic and kinetic energy densities may be, after all, comparable. 
Equipartition is not an experimental
evidence, it is a working hypothesis which may or may not be 
realized in the system under observation. For instance equipartition 
probably holds for the Milky Way but it does not seem to be valid in 
the Magellanic Clouds \cite{magcl}. The average equipartition 
field strengths in galaxies ranges from the $4 \mu$ G of M33 up to the 
$19 \mu$ G of NGC2276 \cite{eq}.

In order to infer the magnitude of the 
magnetic field strength Faraday effect has been 
widely used. When a polarized radio wave passes 
through a region of space containing a plasma with a magnetic field 
the polarization plane of the wave gets rotated by an amount 
which is directly proportional to the square of the plasma 
frequency (and hence to the electron density) and to the 
Larmor frequency (and hence to the magnetic field intensity). 
Calling $ \phi$ the shift in the polarization plane of the 
wave, a linear regression, connecting the shift in the polarization
plane and the square of the wavelength of observation, can be obtained:
\begin{equation}
\phi = \phi_{0} + {\rm RM} \lambda^2.
\label{regr}
\end{equation}
By measuring this relation for two (or more) separate (but close) 
wavelengths, the angular coefficient of the regression can be 
obtained and it turns out to be
\begin{equation}
\frac{ \Delta \phi}{\Delta \lambda^2} = 811.9 \int \biggl(\frac{n_{e}}{
{\rm cm^{-3}}}\biggr) \biggl( \frac{ B_{\parallel}}{\mu {\rm G}}\biggr) d 
\biggl(\frac{\ell}{ {\rm kpc}}\biggr), 
\label{FR}
\end{equation}
in units of ${\rm rad}/{\rm m}^2$ when all the quantities 
of the integrand are measured in the above units. 
As reminded, RM should be performed 
at sufficiently close wave-lengths. Typically the angles should be 
determined with an accuracy greater than $\delta \phi \sim \pm \pi$. Otherwise
ambiguities may arise in the determination of the angular 
coefficient appearing in the linear regression of Eq. (\ref{regr}) 
\cite{rev1,rev3}. 

It should be appreciated that the RM contains 
not only the magnetic field (which should be 
observationally estimated), but also the column density 
of electrons. From the radio-astronomical 
observations, different techniques can be used in order 
to determine the column density of electrons. One 
possibility is to notice that in 
 the observed Universe there 
are pulsars. Pulsars are astrophysical objects emitting 
regular pulses of electromagnetic radiation with periods 
ranging from few milliseconds to few seconds. By comparing the arrival 
times of different radio pulses at different radio wave-lengths, it is 
found that signals are slightly delayed as they pass through the 
interstellar medium exactly because electromagnetic waves travel 
faster in the vacuum than in an ionized medium. Hence, from pulsars 
the column density of electrons can be obtained in the 
form of the  dispersion measure, i.e. 
$DM \propto \int n_{e} d\ell$. Dividing the RM by DM,
an estimate of the magnetic field can be obtained.

Already this simple-minded account of the main experimental techniques 
used for the detection of large scales magnetic fields shows that 
there may be problems in the determination of magnetic 
fields right outside galaxies. There magnetic fields 
are {\em assumed} to be often of ${\rm n G}$ strength. 
However, due to the lack of sources for the determination 
of the column density of electrons, it is hard 
to turn the assumption into an experimental evidence.

Finally, an interesting source of observational informations 
on galactic magnetic field structure is the polarization of 
synchrotron emission. Magnetic fields in external galaxies 
have an uniform and a random component. Synchrotron 
polarization at high radio frequencies (where 
Faraday rotation is small) can be used in order to estimate 
the relative weight of the mean and random parts of a given field since 
the polarization essentially depends upon the ratio between the 
uniform and the total (i.e. random plus uniform) magnetic field \cite{rev7}.

Since various  theoretical speculations  
suggest that also  clusters are magnetized, it would be interesting 
to know if regular Abell clusters posess large scale magnetic fields. 
Different results in this direction have been reported 
\cite{cl1,cl2,cl3,cl4}. Some studies during the past
decade \cite{cl1,cl2}
dealt mainly with the case of a single cluster (more specifically the 
Coma cluster). The idea was to target (with Faraday rotation 
measurements)  radio sources inside the
cluster. However, it was also soon realized that 
the study of many radio sources inside 
different clusters may lead to experimental 
problems due to the sensitivity limitations of 
radio-astronomical facilities. The strategy 
is currently to study a sample of clusters each with one or two 
bright radio-sources inside.

In the past it was shown that regular clusters have  cores with
a detectable component of RM \cite{cl3,cl4}. Recent results 
suggest that $\mu$ Gauss magnetic fields are indeed detected 
inside regular clusters \cite{cl5}. Inside the cluster means 
in the intra-cluster medium. Therefore, these magnetic fields
cannot be associated with individual galaxies.
 
Regular Abell clusters with strong x-ray emission were 
studied using a twofold technique \cite{cl5,cl6}. From the ROSAT 
\footnote{The ROetgen SATellite was flying from June 1991 to February 1999.
ROSAT provided a map of the x-ray sky in the range $0.1$--$2.5$ keV. For the 
ROSAT catalog of X-ray bright Abell clusters see \cite{roscat}.}
\cite{tru}
full sky survey, the electron density has been determined.
Faraday RM (for the same set 
of 16 Abell clusters) has been estimated through observations at the VLA 
\footnote{The Very Large Array
telescope is a radio-astronomical facility consisting 
of 27 parabolic antennas spread  $20 {\rm km}^2$ in the New Mexico desert.}.
The amusing result (confirming previous claims based only on one cluster 
\cite{cl1,cl2}) is that x-ray bright Abell clusters
 possess a magnetic field of $\mu$ Gauss 
strength.The clusters have been selected in order 
to show similar morphological features. All the 16 clusters 
monitored with this technique are at low red-shift ($z<0.1$) 
and at high galactic latitude ($|b|>20^{0}$).

These recent developments are rather promising 
and establish a clear connection between radio-astronomical 
techniques  and the improvements in the knowledge 
of  x-ray sky. There are 
various satellite missions mapping the
x-ray sky at low energies (ASCA, CHANDRA, NEWTON 
\footnote{ASCA is operating between $0.4$ AND $10$ keV and it is 
flying since February 1993. CHANDRA (NASA mission)  and NEWTON (ESA mission) 
have an energy range 
comparable with the one of ASCA and were launched, almost 
simultaneously, in 1999.}). There is 
the hope that a more precise knowledge of the surface brightness of regular
clusters will help in the experimental determination of large scale 
magnetic fields between galaxies.

\begin{figure}[htb]
    \centering
    \includegraphics[height=2.5in]{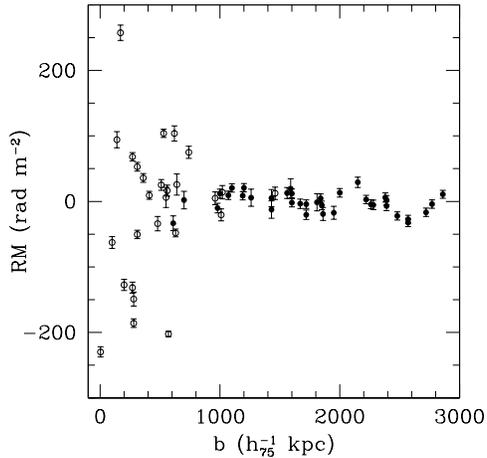}
    \caption{From Ref. \cite{cl5} the RM deduced from a 
sample of 16 X-ray bright Abell clusters is reported as a function 
of the source impact parameter.}
    \label{F1}
\end{figure}

It is interesting to notice that intra-cluster magnetic fields 
of $\mu$ G strength can induce Faraday rotation on CMB polarization.
By combining informations from Sunyaev-Zeldovich effect and X-ray 
emission from the same clusters, it has been recently 
suggested that a richer information concerning electron column 
density can be obtained \cite{ohno}. 
In Fig. \ref{F1} the results reported in \cite{cl5} 
are summarized. In Fig. \ref{F1} the RM 
of the sample of x-ray bright Abell clusters is reported after 
the subtraction of the RM of the galaxy. At high 
galactic latitude (where all the observed clusters are) 
the galactic contribution is rather small and of the 
order of $9.5 {\rm rad}/{\rm m}^2$. 
In Fig. \ref{F1} 
the open points represent sources viewed through 
the thermal cluster gas, whereas the full points 
represent control sources at impact parameters larger than the cluster gas.
The excess in RM attributed to clusters sources is clearly visible.

Using the described techniques large scale magnetic fields can be
observed and studied in external galaxies, in clusters and also in our own 
galaxy.  While the study of external galaxies 
and clusters may provide a global picture of magnetic fields, 
the  galactic observations performed within the Milky Way 
are more sensitive to the local spatial variations of the magnetic field.
For this reasons local and global observations are complementary. The flipped
side of the coin, as we will see in the second part of the present Section, 
is that the global structure of the magnetic field of our galaxy 
is not known directly and  to high precision but it is deduced from 
(or corroborated by) 
 the global knowledge of other spiral galaxies.

\subsection{Geometrical models}
Since the early seventies \cite{rev1,mw1} the magnetic field 
of the Milky way was shown to be parallel to the galactic plane.
RM derived from pulsars allow  consistent determinations 
of the magnetic field direction and intensity \cite{mw2,mw3}. In the Milky Way,
the uniform component of the 
 magnetic field is thought to lie in the plane of 
the galactic disk and it is thought to be directed approximately along the spiral 
arm. There is, though, a slight difference between 
the northern and southern hemisphere. While in the southern hemisphere 
the magnetic field is roughly $2 \mu$G, the  magnetic field 
in the northern emisphere is three times smaller than the 
one of the southern hemisphere. Differently from 
other spirals, the Milky Way has also a large 
radio halo. Finally RM data seem to suggest that the magnetic field 
flips its direction from one spiral arm to the other. 
As far as the stochastic component of the galactic magnetic field 
is concerned, the situation seems to be, according to the reported results,
still unclear \cite{fl1,fl2,fl3}. It is, at present, unclear if the stochastic 
component of the galactic magnetic field is much smaller than 
(or of the same order of)
the related
homogeneous part.  
The geometries of large scale magnetic fields in spiral 
galaxies can be divided into two classes. We can have axysymmetric 
spirals (ASS) and bisymmetric spirals (BSS). This 
classification scheme refers, respectively, to even and odd
parity with respect to rotation by an angle $\pi$ around the 
galactic center. Speculations concerning the origin 
of galactic fields prefer to associate a primordial field with a BSS
configuration while the ASS would be more associated with a field
produced through a strong dynamo activity.

In the case of the Milky way, as we saw, the magnetic field flips 
its direction from one spiral arm to the other and then, as 
pointed out by Sofue and Fujimoto (SF) \cite{mw3} 
the galactic magnetic field should probably be associated with a BSS 
model. In the SF model the radial and azimuthal components 
of the magnetic field in a bisymmetric logarithmic spiral configuration 
is given through
\begin{eqnarray}
&& B_{r} =  f(r) \cos{\biggl( \theta - \beta \ln{\frac{r}{r_{0}}} 
\biggr)} ~\sin{p}
\nonumber\\
&& B_{\theta} =  f(r) \cos{\biggl( \theta - \beta
\ln{\frac{r}{r_0}}\biggr)} \cos{p}
\end{eqnarray}
where $r_{0}\sim 10.5$ kpc is the galactocentric distance 
of the maximum of the field in our spiral arm, $\beta = 1/\tan{p}$  and 
$p=10^{0}$ is the pitch angle of the spiral. The smooth profile 
$f(r)$ can be chosen in different ways. A motivated choice is 
 \cite{mw4,mw5}
\begin{equation}
f(r) = 3 \frac{r_1}{r} \tanh^3{\biggl(\frac{r}{r_2}\biggr)} \mu {\rm G}
\end{equation}
where $r_{1} = 8.5  $ kpc is the distance of the Sun to the 
galactic center and $r_2 = 2 $ kpc. The original 
model of SF does not have dependence  in the $z$ direction, however, 
the $z$ dependence can be included and also more complicated 
models can be built \cite{mw4}. Typically, along the $z$ axis, 
magnetic fields are exponentially suppressed as $\exp{[-z/z_0]}$ 
with $z_{0} \sim 4$ kpc.
The structure of magnetic fields can be relevant when investigating the 
propagation of high-energy protons \cite{bier,far} as noticed already long ago \cite{wolf} 
(see also \cite{mw2,mw3,mw4}).

\renewcommand{\theequation}{3.\arabic{equation}}
\setcounter{equation}{0} 
\section{Magnetic field evolution(s)}
The  
galaxy is a gravitationally bound system formed by fluid of charged 
particles  which is globally 
neutral for scales larger than the Debye sphere.
In the interstellar medium, where the electron density 
is approximately $3\times 10^{-2} ~ {\rm cm}^{-3}$, the 
Debye sphere has a radius of roughly $10$ m.
Moreover, the galaxy
is rotating with a typical rotation period of $3\times10^{8}$ yrs. 
Two complementary descriptions of the plasma  can then  be adopted. 
The first possibility is to study full kinetic system 
(the Vlasov-Landau equations \cite{vla,lan}).
The second (complementary) description relies on the 
 magnetohydrodynamical (MHD)
treatment. Among the discussions of 
MHD effects, dynamo theory is particularly 
relevant. A full
account of the various aspects and debates 
concerning the dynamo theory  can be found 
in excellent textbooks \cite{rev1,rev2} and reviews \cite{kul}.
After some elements of the Vlasov-Landau description, the  
the basic ideas concerning dynamo theory will be introduced.

\subsection{Elements of a kinetic discussion}

Already in flat space \cite{lif}, and, a fortiori, 
in curved space \cite{ber}, 
the kinetic approach is important once we deal with 
electric fields dissipation, charge and current density 
fluctuations and, in more general terms, with all the high 
frequency and small length scale phenomena in the plasma \cite{krall,bis}.
The few elements of kinetic description will be given directly 
in curved spaces since they may be relevant for some applications 
\cite{mg2}.

Consider a conformally flat Friedmann-Robertson-Walker (FRW) 
metric written using the  conformal time coordinate 
\begin{equation}
ds^2 = a^2(\eta)[d\eta^2 - d\vec{x}^2].
\label{metric1}
\end{equation}
Furthermore, consider an equilibrium homogeneous and
isotropic conducting plasma, characterized by a distribution function
$f_0(p)$ common for both positively and negatively charged
ultrarelativistic particles (for example, electrons and positrons) .
Suppose now that this plasma is slightly perturbed, so that the
distribution functions are
\begin{equation}
f_{+}(\vec{x}, \vec{p}, \eta) = f_0(p) + 
\delta f_{+}(\vec{x},\vec{p},\eta),
\,\,\,\,\,\,\,\,f_{-}(\vec{x},\vec{p}, \eta) =  
f_0(p) + \delta f_{-}(\vec{x},\vec{p},\eta),
\end{equation}
where $+$ refers to positrons and $-$ to electrons, and $\vec{p}$ is
the conformal momentum.  The Vlasov equation   defining the
curved-space evolution of the perturbed distributions can  be written
as  \cite{mg2}
\begin{eqnarray} 
&&\frac{\partial f_{+}}{\partial \eta} + 
\vec{v} \cdot \frac{\partial f_{+}}{\partial\vec{x}} + e ( \vec{E} + 
\vec{v}\times \vec{B}) \cdot\frac{\partial f_{+}}{\partial \vec{p}} = 
\biggl( 
\frac{\partial f_{+}}{\partial \eta}\biggr)_{\rm coll}
\label{Vl+},\\
&&\frac{\partial f_{-}}{\partial \eta} + 
\vec{v} \cdot \frac{\partial f_{-}}{\partial\vec{x}} - e ( \vec{E} + 
\vec{v}\times \vec{B})\cdot \frac{\partial f_{-}}{\partial \vec{p}} = 
\biggl(\frac{\partial f_{-}}{\partial\eta}\biggr)_{\rm coll}
\label{Vl-},
\end{eqnarray}
where the two terms appearing at the right hand side of each 
equation are the collision terms. The electric and 
magnetic fields are rescaled by the second power of the scale factor. 
 This system of equation represents 
the curved space extension of the Vlasov-Landau approach  to plasma
fluctuations \cite{vla,lan}. All particle number densities here are
related to the comoving volume.
By subtracting Eqs. (\ref{Vl+}) and (\ref{Vl-})  we obtain the
equations relating the fluctuations of  the distributions functions
of the charged particles present in the plasma  to the induced gauge
field fluctuations:
\begin{eqnarray}
&&\frac{\partial}{\partial\eta} f(\vec{x}, \vec{p},t) + 
\vec{v}\cdot \frac{\partial }{\partial\vec{x}} f(\vec{x},\vec{p},t) 
+ 2 e \vec{E}\cdot 
\frac{\partial f_0}{\partial \vec{p}} =- \nu(p) f,
\nonumber\\
&& \vec{\nabla} \cdot \vec{E} = e \int d^3 p f(\vec{x},\vec{p},\eta),
\nonumber\\
&& \vec{\nabla}\times \vec{E} + \vec{B}' =0,
\nonumber\\
&& \vec{\nabla}\cdot \vec{B} =0,
\nonumber\\
&& \vec{\nabla}\times \vec{B} -\vec{E}'= \int d^3 p \vec{v} 
f(\vec{x}, \vec{p},\eta),
\label{Vlasov}
\end{eqnarray}
where $f(\vec{x}, \vec{p}, \eta) = \delta f_{+}(\vec{x},\vec{p},\eta)
-  \delta f_{-}(\vec{x},\vec{p},\eta)$ and $\nu(p)$ is a typical
frequency of collisions \cite{ber}. 

Now, if $\delta f_{\pm}(\vec{x},\vec{p},\eta)\neq 0$ at the beginning of the
radiation dominated epoch $\eta_0$  and
$E(\vec{x},\eta_0)\simeq B(\vec{x},\eta_0)=0$ initially, the magnetic
field at later times can be found from Eqs. (\ref{Vlasov}) \cite{lif}.
Various useful generalizations of the Vlasov-Landau system to curved spaces 
is given in \cite{vlcur1,vlcur2,vlcur3}.

\subsection{Effective (MHD) description and dynamo instability}

For scales sufficiently large compared with the Debye sphere and for 
frequencies sufficiently small compared with the plasma frequency, the
spectrum of plasma excitations obtained from the kinetic 
theory matches the  spectrum obtained from an (effective) MHD
description \cite{krall}. 
Furthermore, since the galaxy is rotating and since the conditions 
of validity of the MHD approximation
are met, it is possible to use the so-called dynamo instability in order 
amplify a small magnetic inhomogeneity up to the 
observed value. This is, at least, the hope \cite{kul}.

The pioneering attempts towards a MHD description of interstellar plasma, 
go back to the works of Alfv\'en \cite{alv1} and of 
Fermi and Chandrasekar \cite{fermi2}. Since the work of Parker \cite{park},
on the so-called $\alpha-\Omega$ theory, the dynamo effect has been used 
in order to explain (or, to ease) the problem of the origin of 
galactic magnetic fields. The standard dynamo theory 
has been questioned in different ways.  Piddington \cite{pid1,pid2} 
pointed out that small-scale magnetic fields can grow large 
enough (until equipartition is reached) to swamp the dynamo action.
The quenching of the dynamo action has been numerically shown by
Kulsrud and Anderson \cite{ka}. More recently, it has been argued that 
if the large-scale magnetic field reaches the  critical value 
$R{\rm e}_{\rm M}^{-1/2} v$  \footnote{ ${\rm R e}_{M}$ is the magnetic Reynolds 
number \cite{bis}, i.e. , approximately, the ratio of the 
first (the dynamo contribution)  over the second 
term (the magnetic diffusivity term)
 appearing at the right hand side of Eq. (\ref{mdiff}); $v$ is the velocity
field at the outer scale of turbulence.}
the dynamo action could also be quenched \cite{VC1,VC2}.

MHD equations can be derived from a microscopic (kinetic) 
approach and also from a macroscopic approach where 
the displacement current is neglected \cite{krall}. 
If the displacement 
current is neglected the electric field can be expressed 
using the Ohm law and the magnetic diffusivity equation is obtained
\begin{equation}
\frac{\partial \vec{B}}{\partial\eta} = \vec{\nabla} \times(\vec{V} 
\times \vec{B}) + \frac{1}{\sigma} 
\nabla^2 \vec{B}.
\label{mdiff}
\end{equation}
The first term at the r.h.s. of Eq. (\ref{mdiff}) 
is related to the dynamo term, while the second term is the 
diffusivity term.
The conductivity $\sigma$ appearing in Eq. (\ref{mdiff})
is a global quantity which can be computed in a kinetic approach \cite{krall}
during a given phase of evolution of the background 
geometry \cite{mg2}. If the plasma is non-relativistic $\sigma \propto T^{3/2}$.
If the plasma is relativistic $\sigma \propto T$. 
In Eq. (\ref{mdiff}) 
the contribution containing the conductivity 
is usually called magnetic diffusivity term. The magnetic 
diffusivity scale is 
defined as 
\begin{equation}
L_{\sigma} = \sqrt{\frac{\tau_{U}}{\sigma}},
\end{equation}
where $\tau_{U}$ is the age of the Universe at the corresponding epoch.
Typical galactic values are $1/\sigma \sim 10^{25} {\rm cm}^2/{\rm sec}$, 
$H_{0} \sim \tau_{U}^{-1} \sim 10^{-18} $ Hz, $L_{\sigma } \sim {\rm A.U.}$ . 

Eq. (\ref{mdiff}) is exact, in the sense 
that both $\vec{V}$ and $\vec{B}$ contain long and short 
wavelength modes. The aim of the 
various attempts of the dynamo theory is to get 
an equation describing only the ``mean value'' of the magnetic field.
To this end the fisrt step is to separate 
the exact magnetic and velocity fields as 
\begin{eqnarray}
&&\vec{B} = \langle \vec{B} \rangle + \vec{b},
\nonumber\\
&& \vec{V} = \langle \vec{V} \rangle + \vec{v},
\label{sep}
\end{eqnarray}
where $\langle \vec{B} \rangle$ and  $\langle \vec{V} \rangle$
are the  averages over an ensemble of many realizations 
of the velocity field $\vec{V}$. 
In order to derive the standard form of the dynamo equations
 few important assumptions should be 
made. These assumptions are :
\begin{itemize}
\item{-} The scale of variation of the turbulent motion $\vec{v}$ 
should be smaller than the typical scale of variation of $\langle \vec{B}\rangle$.
In the galactic problem $\langle \vec{V}\rangle$ is the differential rotation of the 
galaxy, while $\vec{v}$ is the turbulent motion generated by 
stars and supernovae. Typically the scale of variation of $\vec{v}$ is 
less than $100$ pc while the interesting scales for $\langle \vec{B} \rangle$ 
are  larger than the kpc.

\item{-} The field $\vec{b}$ is such that $|\vec{b}|\ll |\langle \vec{B} \rangle|$.

\item{-} It should happen 
that $\langle \vec{v} \cdot \vec{\nabla} \times \vec{v} \rangle \neq 0$. 

\item{-} Magnetic flux is frozen into the plasma 
(i.e. magnetic flux is conserved).
\end{itemize}
From the magnetic diffusivity equation (\ref{mdiff}), and using 
the listed assumptions, it is possible to
derive the typical structure of the dynamo term by carefully averaging
over the velocity field according to the procedure outlined in 
\cite{rev1,rev2,kul}. Inserting Eq. (\ref{sep}) into (\ref{mdiff}) 
and breaking the equation into a mean part and a random part,
two separate induction equations can be obtained 
for the mean and random parts of the magnetic field
\begin{eqnarray}
&&\frac{\partial \langle \vec{B} \rangle}{\partial\eta} = 
\vec{\nabla} \times \biggl( \langle \vec{V}\rangle \times \langle 
\vec{B} \rangle\biggr) + 
\vec{\nabla} \times \langle 
\vec{v} \times \vec{b} \rangle,
\label{mean}\\
&& \frac{\partial \vec{b}}{\partial\eta} = \vec{\nabla} \times ( \vec{v} \times \langle \vec{B}\rangle) +
\vec{\nabla} \times( \langle \vec{V} \rangle \times \vec{b}) + \vec{\nabla}\times( \vec{v} \times \vec{b} ) -
 \vec{\nabla} \times \langle 
\vec{v} \times \vec{b} \rangle,
\label{random} 
\end{eqnarray}
where the (magnetic) diffusivity terms have been neglected.
In Eq. (\ref{mean}), $ \langle \vec{v} \times \vec{b} \rangle$ is called
``turbulent emf'' and it is the average of the cross 
product of the small-scale velocity field $\vec{v}$ and of the small 
scale magnetic field $\vec{b}$ over a scale much smaller than the scale 
of $\langle \vec{B} \rangle$ but much larger than the scale of turbulence.
Sometimes, the calculation of the effect of $\langle \vec{v} \times \vec{b}\rangle$ 
is done in the case of incompressible and isotropic turbulence. In this case 
$\langle \vec{v} \times \vec{b}\rangle =0$. This estimate is, however, not realistic 
since $\langle \vec{B}\rangle$ is not isotropic. 
More correctly \cite{kul}, $\langle \vec{v} \times \vec{b}\rangle$ should be 
evaluated by using Eq. (\ref{random}) which is usually written in a simplified form 
\begin{equation}
 \frac{\partial \vec{b}}{\partial\eta} =    \vec{\nabla} \times ( \vec{v} \times \langle \vec{B}\rangle),
\label{inran}
\end{equation}
where all but the first term of Eq. (\ref{random}) have been 
neglected. To neglect the term $ \vec{\nabla} \times (\langle\vec{V}\rangle \times \vec{b})$ 
does not pose any problem since it corresponds to choose a reference 
frame where $\langle \vec{V} \rangle$ is constant. 
However, the other terms, neglected in Eq. (\ref{inran}), 
are dropped because it is assumed that $|\vec{b}| \ll |\langle \vec{B}\rangle| $. 
This assumption may not be valid all the time and for all the scales. 
The validity of Eq. (\ref{inran}) seems to require that $1/\sigma$ is very large 
so that magnetic diffusivity can keep always $\vec{b}$ small \cite{KR}.
On the other hand \cite{kul} one can argue that $\vec{b}$ is only present 
over very small scales (smaller than $100$ pc) and in this case 
the approximate form of eq. (\ref{inran}) seems to be more justified.

From Eqs. (\ref{mean})--(\ref{inran}) 
 it is possible to get to the final result 
for the evolution equation of $\langle \vec{B} \rangle$ \cite{kul} as it
is usally quoted
\begin{equation}
\frac{\partial \langle \vec{B}\rangle}{\partial\eta} =
\vec{\nabla}\times (\alpha\langle \vec{B}\rangle) +
\beta\nabla^2\langle\vec{B}\rangle + \vec{\nabla} \times \biggl( \langle \vec{V} \rangle \times 
\langle \vec{B}\rangle \biggr), 
\label{fin}
\end{equation}
where 
\begin{eqnarray}
&&\alpha 
= -\frac{\tau_{0}}{3}\langle\vec{v}\cdot\vec{\nabla}
\times\vec{v}\rangle,
\label{at}\\
&& \beta = \frac{\tau_0}{3} \langle \vec{v}^2\rangle,
\label{bt}\end{eqnarray}
where $\alpha$ is the dynamo term, $\beta $ is the diffusion term 
and $\tau$ is the typical correlation time of the velocity field.
The term $\alpha$ is, in general, space-dependent.

It is  interesting to point out \cite{rev2} that the
dynamo term in Eq. (\ref{fin}) has a simple electrodynamical meaning (when 
$\alpha$ is constant),
namely, it can be interpreted as a mean ohmic current directed along
the magnetic field:
\begin{equation}
\vec{J} = - \alpha \vec{B}.
\label{2.10}
\end{equation}
This equation tells us that an ensemble of screw-like vortices with
zero mean helicity is able to generate loops in the magnetic flux
tubes in a plane orthogonal to the one of the original field.

If the velocity field {\em is}
parity-invariant (i.e. no vorticity for scales comparable with the
correlation length of the magnetic field), then the dynamics of the
infrared modes is decoupled from the velocity field since, over those
scales, $\alpha =0$. When the (averaged) dynamo term dominates in 
Eq. (\ref{fin}), magnetic fields can be exponentially amplified. 
The standard lore is that the dynamo action stops when 
the value of the magnetic field reaches the equipartition value 
(i.e. when the magnetic and kinetic energy of the plasma are comparable).
At this 
point the dynamo ``saturates''. This statement means that, in more 
 dynamical terms
\cite{rev1}, back-reaction effects cannot be neglected anymore and 
Eq. (\ref{fin}) should then be 
supplemented with non-linear terms (of the order of $\vec{B}^2$),
 whose effect is to stabilize
the amplification of the magnetic field.

It is sometimes useful to recall that the full 
MHD equations can be studied in two different limits : the ideal (or 
superconducting) approximation where the conductivity is 
assumed to be very high and the real (or resistive) limit 
where the conductivity takes a finite value. 
In the ideal limit both the magnetic flux and the magnetic helicity 
are conserved. This means, formally \cite{mg1}, 
\begin{eqnarray}
\frac{d}{d\eta} \int_{\Sigma} \vec{B} \cdot d\vec{\Sigma}=-
\frac{1}{\sigma} \int_{\Sigma} \vec{\nabla} \times\vec{\nabla}
\times\vec{B}\cdot d\vec{\Sigma},
\label{flux}
\end{eqnarray}
where $\Sigma$ is an arbitrary closed surface which moves with the
plasma.
If we are in the in the inertial regime (i.e. $L>L_{\sigma}$) we can
say that the expression appearing at the right hand side is
sub-leading
and the magnetic flux lines evolve glued to the plasma element.

The other quantity which is conserved in the superconducting limit
is the magnetic helicity
\begin{equation}
{\cal H}_{M} = \int_{V} d^3 x \vec{A}~\cdot \vec{B},
\label{h1}
\end{equation}
where $\vec{A}$ is the vector potential \footnote{Notice that in conformally
flat FRW spaces the radiation gauge is conformally invariant. This property
is not shared by the Lorentz gauge condition \cite{ford}.}.
In Eq. (\ref{h1}) the vector potential appears and, therefore
it might seem that the expression is not gauge invariant. This is not
the case. In fact $\vec{A}\cdot\vec{B}$ is not gauge invariant but,
none the less, ${\cal H}_{M}$ is gauge-invariant since 
the integration volume is defined in such a way that the magnetic field
$\vec{B}$
is parallel to the surface which bounds $V$ and which we will call
$\partial V$. If $\vec{n}$ is the unit vector normal to $\partial V$
then $\vec{B}\cdot\vec{n}=0$ on $\partial V$ \cite{mmm}. 

The magnetic gyrotropy
\begin{equation}
\vec{B}\cdot\vec{\nabla} \times\vec{B}
\end{equation}
it is a gauge invariant measure of the diffusion rate of ${\cal H}_{M}$
at finite conductivity. In fact \cite{mg1}
\begin{equation}
\frac{d}{d\eta} {\cal H}_{M} = - \frac{1}{\sigma} \int_{V} d^3 x
{}~\vec{B}\cdot\vec{\nabla} \times\vec{B}.
\label{h2}
\end{equation}
The magnetic gyrotropy is a useful quantity in order to distinguish 
different mechanisms for the magnetic field generation. Some 
mechanisms are able to produce magnetic fields whose 
flux lines have a topologically non-trivial structure (i.e. 
$\langle \vec{B} \cdot \vec{\nabla} \times \vec{B} \rangle \neq 0$).

Usually the picture 
for the formation of galactic magnetic fields is related to the 
possibility of implementing the dynamo mechanism.  
By comparing 
the rotation period with the age of the galaxy (for a Universe with 
$\Omega_{\Lambda} \sim 
0.7$, $h \sim 0.65$ and $\Omega_{\rm m} \sim 0.3$) the number of rotations
performed by the galaxy since its origin is approximately  $30$. 
During these $30$ rotations the dynamo term of Eq. (\ref{fin}) 
 dominates against the magnetic diffusivity. As a 
consequence an instability develops. This instability can be used
in order to drive the magnetic field from some small initial condition
up to its observed value.
Eq. (\ref{fin}) is linear in the mean 
magnetic field. Hence, initial conditions for the mean magnetic field
 should be postulated at a given time and over a given scale. 
This initial mean field, postulated as initial 
condition of (\ref{fin}) is usually called seed. 

Most of the work in the context of the dynamo 
theory focuses on reproducing the correct features of the 
magnetic field of our galaxy.
The achievable amplification produced by the 
dynamo instability can be at most of $10^{13}$, i.e. $e^{30}$. Thus, if 
the present value of the galactic magnetic field is $10^{-6}$ Gauss, its value 
right after the gravitational collapse of the protogalaxy might have 
been as small as $10^{-19}$ Gauss over a typical scale of $30$--$100$ kpc.

There is a simple way to relate the value of the magnetic fields 
right after gravitational collapse to the value of the magnetic field 
right before gravitational collapse. Since the gravitational collapse 
occurs at high conductivity the magnetic flux and the magnetic helicity
are both conserved. Right before the formation of the galaxy a patch 
of matter of roughly $1$ Mpc collapses by gravitational 
instability. Right before the collapse the mean energy density  
of the patch, stored in matter, 
 is of the order of the critical density of the Universe. 
Right after collapse the mean matter density of the protogalaxy
is, approximately, six orders of magnitude larger than the critical density.

Since the physical size of the patch decreases from $1$ Mpc to 
$30$ kpc the magnetic field increases, because of flux conservation, 
of a factor $(\rho_{\rm a}/\rho_{\rm b})^{2/3} \sim 10^{4}$ 
where $\rho_{\rm a}$ and $\rho_{\rm b}$ are, respectively the energy densities 
right after and right before gravitational collapse. The 
correct initial condition in order to turn on the dynamo instability
is $B \sim 10^{-23}$ Gauss over a scale of $1$ Mpc, right before 
gravitational collapse. 

Since the flux is conserved the ratio between the magnetic energy 
density, $\rho_{\rm B}(L,\eta)$ 
 and the energy density sitting in radiation, $\rho_{\gamma}(\eta)$
 is almost constant and therefore, in terms of this quantity (which is only scale 
dependent but not time dependent), the dynamo requirement can be rephrased as
\begin{equation}
r_{\rm B}(L) = \frac{\rho_{\rm B}(L,\eta)}{\rho_{\gamma}(\eta)} 
\geq 10^{-34},\,\,\,\, L\sim 1\,{\rm Mpc},
\label{dyn}
\end{equation}
to be compared with the value $r_{\rm B} \sim 10^{-8}$ which would lead to
the galactic magnetic field only thanks to the collapse 
and without the need of dynamo action. This is 
the case when the magnetic field is fully primordial.

The estimate of Eq. (\ref{dyn})  is, to say the least, 
rather generous and has been presented just in order to 
make contact  with several papers (concerned with the origin 
of large scale magnetic fields) using such an estimate.
Eq. (\ref{dyn})  is based on the (highly questionable) 
assumption that the amplification 
occurs over thirty e-folds while the magnetic flux is 
completely frozen in. In the real situation, the 
achievable amplification is much smaller. Typically a good 
seed would not be $10^{-19}$ G after collapse (as we assumed for 
the simplicity of the discussion) but of the order of (or larger 
than) $10^{-12}$--$10^{-13}$ G \cite{kul}. 

The possible applications of dynamo mechanism to  clusters is still
under debate and it seems more problematic \cite{cl5,cl6,dc}.  
The typical scale of the gravitational collapse of a cluster 
is larger (roughly by one order of magnitude) than the scale of gravitational
collapse of the protogalaxy. Furthermore, the mean mass density 
within the Abell radius ( $\simeq 1.5 h^{-1} $ Mpc) is roughly 
$10^{3}$ larger than the critical density. Consequently, clusters 
rotate less than galaxies since their origin and the value of 
$r_{\rm B}(L)$ has to be larger than in the case of galaxies. 
Since the details of the dynamo mechanism applied to clusters are 
not clear, at present, it will be required that $r_{\rm B}(L_{\rm Mpc}) \gg 10^{-34}$
[for instance $r_{\rm B} (L_{\rm Mpc}) \simeq 10^{-12}$]. 

\renewcommand{\theequation}{4.\arabic{equation}}
\setcounter{equation}{0} 
\section{A twofold path to the origin}
Back in the late sixties harrison
\cite{second} suggested that the initial conditions 
of the magnetic diffusivity equation might have something 
to do with cosmology in the same way as 
he  suggested that the primordial spectrum of gravitational potential 
fluctuations (i.e. the Harrison-Zeldovich spectrum) 
might be produced in some primordial phase of the evolution of the Universe.
Since then, several mechanisms have been invoked 
in order to explain the origin of the magnetic seeds \cite{s1,s3} and few of them 
are compatible with inflationary evolution. It is not my purpose 
to review here all the different mechanisms which have been proposed and 
good reviews exist already \cite{en,rub,dol}. 
A very incomplete 
selection of references is, however, reported \cite{s1,s2,s3}.  
Furthermore, more details on this topic can be found in the 
contribution of D. Boyanovsky \cite{db}.

In spite of the richness of the theoretical 
models, the mechanisms for magnetic field generation can be divided,
broadly speaking, into two categories: astrophysical \cite{rev3,kul} and 
cosmological. The cosmological mechanisms can be divided, in their turn,
into {\em causal} mechanisms (where the magnetic seeds are produced at a given time inside the 
horizon) and {\em inflationary} mechanisms where correlations in the magnetic field 
are produced outside the horizon. Astrophysical mechanisms 
have always to explain the initial conditions of Eq. (\ref{fin}). This is 
because the MHD are linear in the magnetic fields. It is 
questionable if  purely astrophysical considerations 
can set a natural initial condition for the dynamo amplification.

\subsection{Turbulence?}
Causal mechanisms usually fail 
in reproducing the correct correlation scale of the field whereas 
inflationary mechanisms have problems in reproducing the correct amplitude
required in order to turn on successfully the dynamo action.
In the context of causal mechanisms there are interesting 
proposals in order to enlarge the correlation scale. These 
proposals have to do with the possible occurrence of turbulence 
in the early Universe.  
The ratio of the magnetic Reynolds number 
to the kinetic Reynolds number is the Prandtl number \cite{bis}
\begin{equation}
{\rm Pr}_{\rm M} = \frac{{\rm Re}_{\rm M}}{{\rm Re}} = \nu \sigma, 
\label{pr}
\end{equation}
where $\nu$ is the thermal diffusivity coefficient 
and $\sigma$, is, as usual, the conductivity.
Consider, for instance, the case of the electroweak epoch 
\cite{turb1,turb2,turb3,mg3,mg4,turb4}. At this epoch 
taking $H_{\rm ew}^{-1} \sim 3 {\rm cm}$ we get that 
${\rm Pr}_{\rm M}\sim \nu\sigma \sim 10^{6} $ where the bulk velocity 
of the plasma is of the order of the bubble wall velocity 
at the epoch of the phase transition. 

This means that the early universe is both kinetically and magnetically 
turbulent. The features of magnetic and kinetic 
turbulence are different. This aspect reflects in 
a spectrum of fluctuations is different from the usual 
Kolmogorov spectrum \cite{bis}. If the Universe is both 
magnetically and kinetically turbulent it has been speculated 
that an inverse cascade mechanism can occur 
\cite{turb1,turb2,turb3,turb4}. This idea was originally 
put forward in the context of MHD simulations \cite{bis}. 
The inverse cascade would imply a growth in the correlation scale 
of the magnetic inhomogeneities and it has been shown to 
occur numerically in the approximation of unitary 
Prandtl number \cite{bis}. Specific cascade models 
have been also studied \cite{turb1,turb2,turb3,turb5a}. 
A particularly important r\^ole is played, in this context, 
by the initial spectrum of magnetic fields (the so-called 
injection spectrum \cite{turb3}) and by the topological 
properties of the magnetic flux lines. If the 
system has non vanishing magnetic helicity and magnetic gyrotropy
it was suggested that the inverse cascade can occur more efficiently 
\cite{turb5,turb6}. Recently simulations have discussed 
the possibility of inverse cascade in realistic MHD models \cite{turb6}. 
More analytic discussions based on renormalization group 
approach applied to turbulent MHD seem to be not totally consistent with
the occurrence of inverse cascade at large scales \cite{bere}.

\subsection{Magnetic fields from dynamical gauge couplings}

Large scale magnetic fluctuations can be generated 
during the early history of the Universe and can go 
outside the horizon with a mechanism similar 
to the one required in order to produce fluctuations 
in the gravitational (Bardeen) potential. In this case the correlation 
scale of the magnetic inhomogeneities can be large. However, the 
typical amplitudes obtainable in this class of models may be too small.

The key property allowing the amplification of the fluctuations 
of the scalar and tensor modes of the geometry is the fact 
that the corresponding equations of motion are not invariant 
under Weyl rescaling of a (conformally flat) metric of FRW type.
In this sense the evolution equations of relic gravitons 
and of the scalar modes of the geometry are  said to be {\em 
not conformally 
invariant}. If gauge couplings are dynamical, the evolution 
equations of the gauge field are also not conformally invariant.
Interesting examples in this direction are models containing 
extra-dimensions and scalar-tensor theories of gravity 
where the gauge coupling is, effectively, a scalar 
degree of freedom evolving in a given geometry.

The remarkable similarity of the 
abundances of light elements in different 
galaxies leads to postulate that the 
Universe had to be dominated by radiation 
at the moment when the light elements were formed, 
namely for temperatures of approximately $0.1 $ MeV 
\cite{sar1,sar2}. 
Prior to the moment of nucleosynthesis 
even indirect informations concerning the thermodynamical 
state of our Universe are lacking even if our knowledge 
of particle physics could give us important hints concerning 
the dynamics of the electroweak phase transition \cite{mis}.
 
The success  of big-bang nucleosynthesis (BBN) 
sets limits on  alternative cosmological scenarios. 
Departures from homogeneity \cite{hom} and isotropy \cite{iso} 
of the background 
geometry can be successfully constrained. In the same spirit,
BBN can also set limits on the 
dynamical evolution of internal dimensions \cite{int1,int2}.
Internal dimensions are an essential ingredient 
of  theories attempting the unification of gravitational and 
gauge interactions in a higher dimensional background like 
Kaluza-Klein theories \cite{kk}  and superstring theories \cite{ss}. 
 
Defining, respectively, $b_{BBN}$ 
and $b_0$ as the size of  the internal dimensions at the BBN 
time and at the present epoch, the maximal variation 
allowed to the internal scale factor from the BBN time 
can be expressed as $b_{BBN}/b_0 \sim 1 
+ \epsilon$ where $ |\epsilon | < 10^{-2} $ \cite{int1,int2}. 
The bounds on the  variation 
of the internal dimensions during the matter dominated epoch
are even stronger. Denoting with an over-dot the derivation with 
respect to the cosmic time coordinate, we have that 
$ |\dot{b}/b| < 10^{-9} H_0$ where 
$H_0$ is the present value of the Hubble parameter \cite{int1}.
The fact that the time evolution of internal dimensions
is so tightly constrained for temperatures lower of $1$ MeV
does not forbid that they could have been dynamical 
prior to that epoch. Moreover, recent observational 
evidence \cite{we1,we2,we3} seem to imply that 
the fine structure constant can be changing even today.

Suppose that prior to BBN internal dimensions 
were evolving in time  and assume, for sake of simplicity, that 
after BBN the internal dimensions have been frozen to their present 
(constant) value. 
Consider a homogeneous and anisotropic manifold 
whose line element can be written as 
\begin{eqnarray}
ds^2 = G_{\mu\nu} dx^{\mu} dx^{\nu} = 
a^2(\eta) [ d\eta^2 - \gamma_{i j} d x^{i} dx^{j}] - b^2(\eta) \gamma_{a b} 
dy^a d y^b,
\nonumber \\
\mu,\nu = 0,..., D-1=d+n , ~~~~~~ i, j=1,..., d , ~~~~~~ 
a,b = d+1,..., d+n.
\label{metric}
\end{eqnarray}
[$\eta$ is the
conformal time coordinate related, as usual to the cosmic time $t=\int a(\eta)
d\eta$ ; $\gamma_{ij}(x)$, $\gamma_{ab}(y)$ are the metric
tensors of two maximally symmetric Euclidean 
manifolds parameterized,
respectively, by the ``internal" and the ``external" coordinates $\{x^i\}$ and
$\{y^a\}$]. 
The metric of Eq. (\ref{metric})
 describes the situation in which the $d$ external dimensions 
(evolving
with scale factor $a(\eta)$) and  the $n$ internal ones (evolving with scale
factor $b(\eta)$) are dynamically decoupled from each other \cite{gio1}. 
The results of the present investigation, however,
can be easily generalized to the case of $n$ different scale factors in the 
internal manifold.

Consider now  a pure electromagnetic fluctuation decoupled 
from the sources, representing an electromagnetic wave propagating 
in the $d$-dimensional external space such that $A_{\mu} \equiv A_{\mu}(\vec{x}, 
\eta)$, 
$A_{a} =0$. In the metric given in Eq. (\ref{metric}) the evolution 
equation of the gauge field fluctuations can be written as 
\begin{equation}
\frac{1}{\sqrt{-G}} \partial_{\mu}\biggl( \sqrt{-G} G^{\alpha\mu} 
G^{\beta\nu} F_{\alpha\beta} \biggr) =0,
\end{equation}
where $F_{\alpha\beta} = \nabla_{[\alpha}A_{\beta]}$ 
is the gauge field strength and 
$G$ is the determinant of the $D$ dimensional metric. Notice that 
if $n=0$ the space-time is isotropic and, therefore, the Maxwell's 
equations can be reduced (by trivial rescaling) to the flat space equations. 
If $n \neq 0$ we have that 
the evolution equation of the electromagnetic fluctuations propagating in the 
external $d$-dimensional manifold will receive a contribution from the internal
 dimensions which cannot be rescaled away. 

In the radiation gauge ($A_0 =0$ and $\nabla_{i} A^{i} =0$) 
the evolution  the vector potentials can be written as 
\begin{equation}
A_{i}'' + n {\cal F} A_{i}' - \vec{\nabla}^2 A_{i} =0, \,\,\,\,\,
 {\cal F} = \frac{b'}{b}.
\label{vec1}
\end{equation}
The vector potentials $A_{i}$ are already rescaled with respect 
to the (conformally flat) $d+1$ dimensional metric.  
In terms of the canonical normal modes of oscillations ${\cal A}_{i} = b^{n/2}
 A_{i}$ 
the previous equation can be written in a simpler form, namely 
\begin{equation}
{\cal A}_{i}''  - V(\eta) {\cal A}_{i} -\vec{\nabla}^2 {\cal A}_{i}  =0,\,\,\,
\,\, 
V(\eta) 
= \frac{n^2}{4} {\cal F}^2 + \frac{n}{2}{\cal F}'.
\label{eq1}
\end{equation}
From this set of equations the induced large scale magnetic fields 
can be computed in various models for the evolution 
of the internal manifold \cite{mg}. It should be 
noticed that large magnetic seeds are produced in this context 
only if internal dimensions are rather 
large if compared to the Planck length. This 
requires a careful discussion of the localization 
properties of gauge fields in the presence 
of large extra-dimensions \cite{loc}  which is beyond 
the scope of the present discussion.

The same effect of magnetic field generation can be also 
obtained in the case of time-evolving gauge coupling 
already in four dimensions. In order to 
emphasize this phenomenon it will now be shown how 
squeezed states of relic photons can be produced \cite{sqback}.
It will be imagined that quantum-mechanical fluctuations 
of the gauge field will be present at some initial 
stage in the evolution of the Universe.

The squeezed states formalism has been successfully
applied to the analysis of tensor,  scalar \cite{gr} 
and rotational \cite{gr2} 
fluctuations of the metric 
by Grishchuk and collaborators. 
 In the case of relic gravitons and relic phonons the analogy with
quantum optics is certainly very inspiring. In the 
case of relic photons the analogy is even closer since 
the time variation of the dilaton coupling plays directly the 
r\^ole of the laser ``pump'' which is employed in order 
to produce experimentally observable squeezed states \cite{sq}. 

The effective 
action of a generic Abelian gauge field in four space-time 
dimensions reads 
\begin{equation}
S = - \frac{1}{4} \int d^4 x \sqrt{- G} \frac{1}{g^2} 
F_{\alpha\beta} ~F^{\alpha\beta},
\label{actionG}
\end{equation}
where $F_{\alpha\beta} = \nabla_{[\alpha} A_{\beta]}$ is the
 Maxwell field strength and $\nabla_{\alpha}$ is the covariant 
derivative with respect to the string frame metric $G_{\mu\nu}$. 
In Eq. (\ref{actionG}) $g$ is the (four dimensional)
gauge coupling which is related to the expectation value 
of a scalar degree of freedom.

From Eq. (\ref{actionG}) it is possible to derive 
the Hamiltonian and the Hamiltonian density 
of the gauge field fluctuations
\begin{eqnarray}
&&H(\eta) = 
\int d^3 k \sum_{\alpha}\biggl[ k~(\hat{a}^{\dagger}_{k,\alpha} 
\hat{a}_{k,\alpha}
+ \hat{a}^{\dagger}_{-k,\alpha} \hat{a}_{-k,\alpha} + 1) 
\nonumber\\
&&+ \epsilon(g) \hat{a}_{-k,\alpha} \hat{a}_{k,\alpha} + \epsilon^{\ast}(g) 
\hat{a}^{\dagger}_{k,\alpha} \hat{a}^{\dagger}_{-k,\alpha}\biggr], ~~~ 
\epsilon(g) = i \frac{g'}{g}.
\end{eqnarray}
where 
The (two-modes) Hamiltonian 
contains  a free part and the effect of the variation 
of the coupling constant is  encoded in the 
(Hermitian) interaction term which is quadratic in the creation 
and annihilation operators whose evolution equations, read,
in the Heisenberg picture
\begin{eqnarray}
&&\frac{ d \hat{a}_{k,\alpha}}{d \eta} = - i k 
\hat{a}_{k,\alpha} - \frac{ g'}{g} 
\hat{a}^{\dagger}_{-k,\alpha},
\nonumber\\
&&
\frac{ d \hat{a}^{\dagger}_{k,\alpha}}{d \eta} = 
i k \hat{a}^{\dagger}_{k,\alpha} - \frac{ g'}{g} 
\hat{a}_{-k,\alpha}.
\label{heiseq}
\end{eqnarray}
The general solution of the previous system of equations can be written 
in terms of  a Bogoliubov-Valatin transformation
\begin{eqnarray}
&& \hat{a}_{k,\alpha}(\eta) = \mu_{k,\alpha}(\eta) \hat{b}_{k,\alpha} + 
\nu_{k,\alpha}(\eta)\hat{b}^{\dagger}_{-k,\alpha}
\nonumber\\
&& \hat{a}^{\dagger}_{k,\alpha}(\eta) 
= \mu^{\ast}_{k,\alpha}(\eta) \hat{b}^{\dagger}_{k,\alpha} + 
\nu_{k,\alpha}^{\ast}(\eta)\hat{b}_{-k,\alpha}
\label{heis}
\end{eqnarray}
where $\hat{a}_{k,\alpha}(0) = 
\hat{b}_{k,\alpha}$ and $\hat{a}_{-k,\alpha}(0) = \hat{b}_{-k,\alpha}$. 
Unitarity requires that  
 the two complex functions $\mu_{k}(\eta)$ and $\nu_{k}(\eta)$ 
are subjected to the condition $|\mu_{k}(\eta)|^2 - |\nu_{k}(\eta)|^2 =1$ 
which also implies that
$\mu_{k}(\eta)$ and $\nu_{k}(\eta)$ can be parameterized in terms of 
one real amplitude and two real phases
\begin{equation}
\mu_k = e^{i \theta_{k}} \cosh{r_{k}},
~~~~\nu_k = e^{ i(2\phi_{k} -  \theta_{k})} 
\sinh{r_{k}},
\label{sq}
\end{equation}
($r$ is sometimes called squeezing parameter and $\phi_{k}$
is the squeezing phase; from now on we will drop the subscript 
labeling each polarization if not strictly necessary).
The total number of produced photons  
\begin{equation}
\langle 0_{-k} 0_{k}| 
\hat{a}^{\dagger}_{k}(\eta) \hat{a}_{k}(\eta) + 
\hat{a}_{-k}^{\dagger} \hat{a}_{-k} |0_{k} 0_{-k}\rangle= 
2 ~\overline{n}_k.
\label{num}
\end{equation}
is expressed in terms of
 $\overline{n}_{k} =\sinh^2{r_{k}}$, i.e. the  mean number 
of produced photon pairs in the mode $k$.
Inserting  Eqs. (\ref{heis}),(\ref{sq}) and (\ref{num}) 
into Eqs. (\ref{heiseq})  
we can derive a closed system involving only the 
$\overline{n}_k$  and the related phases:
\begin{eqnarray}
&&\frac{d  \overline{n}_{k}}{d \eta} = 
-2 f(\overline{n}_{k}) \frac{g'}{g} 
\cos{2 \phi_{k}},
\label{I}\\
&& \frac{d \theta_{k}}{d\eta} = - k + \frac{g'}{g} 
\frac{\overline{n}_{k}}{f(\overline{n}_{k})}
\sin{2 \phi_{k}} ,
\label{II}\\
&& \frac{ d \phi_{k}}{ d \eta} = - k + \frac{g'}{ g} 
\frac{d f(\overline{n}_{k})}{d \overline{n}_{k}}
\sin{ 2 \phi_{k}},
\label{III}
\end{eqnarray}
where $f(\overline{n}_{k}) = \sqrt{ \overline{n}_{k}(\overline{n}_{k} + 1)}$.

The two-point function of the magnetic fields
\begin{equation}
 {\cal G}_{ij} (\vec{r}, \eta) = \langle 0_{-k} 0_{k}|
\hat{B}_{i}(\vec{x}, \eta)
\hat{B}_{j}(\vec{x} + \vec{r},\eta) | 0_{k} 0_{-k}\rangle
\end{equation}
can be expressed, using Eqs. (\ref{heis}) and (\ref{sq}) 
\begin{equation}
{\cal G}_{ij} (\vec{r}) = \int d^3 k 
{\cal G}_{ij}(k) e^{i \vec{k}\cdot\vec{r}}
\end{equation}
where
\begin{eqnarray}
&&{\cal G}_{ij}(k,\eta) = \frac{g^2(\eta) {\cal K}_{i j}}{2(2\pi)^3 a^4(\eta)} 
k[2\sinh^2{r_{k}} +\sinh{2 r_{k}}\cos{ 2\phi_{k}}]
\nonumber\\
&&{\cal K}_{ij} = 
\sum_{\alpha}  e^{\alpha}_{i}(k) e^{\alpha}_{j}(k) = \biggl( \delta_{ij} 
- \frac{k_{i} k_{j}}{k^2}\biggr), 
\label{corr}
\end{eqnarray}
(the vacuum contribution, occurring for $r_{k}=0$, has been 
consistently subtracted).
The intercept for $\vec{r}=0$ of the two-point function traced 
with respect to the two polarizations is related to the magnetic energy density
\begin{equation}
\frac{d \rho_{B}}{d\ln{\omega}} \simeq  
\frac{g^2(\eta)\omega^4}{2\pi^2} 
[2\sinh^2{r_{k}} + \sinh{2 r_{k}}\cos{ 2\phi_{k}}]
\label{endens}
\end{equation}
(where $\omega= k/a$ is the physical frequency).
The two-point function and its trace only 
depend upon $\overline{n}_{k}$ and upon $\phi_{k}$. 
Since 
Eqs. (\ref{I}) and (\ref{III}) do not contain 
any dependence upon $\theta_{k}$ we can attempt to solve the time evolution
by solving them simultaneously. In terms of the new variable 
$x = k \eta$ Eqs. (\ref{I}) and (\ref{III}) can be written as 
\begin{eqnarray}
&&\frac{d \phi_{k}}{d x} = -1 + \frac{d \ln{g}}{d x} 
\frac{d f(\overline{n}_{k})}{d \overline{n}_{k}} \sin{2 \phi_{k}},
\label{IV}\\
&&\frac{d \overline{n}_{k}}{d\ln{g}} = - 2 f(\overline{n}_{k}) 
\cos{2 \phi_{k}},
\label{V}
\end{eqnarray}
If $|(d\ln{g}/dx)(d f(\overline{n}_k)/d \overline{n}_k) 
\sin{2 \phi_{k}}| > 1$, then Eqs. (\ref{IV}) and 
(\ref{V}) can be written as 
\begin{equation}
\frac{d u_{k}}{d \ln{g}} = 
2 \frac{d f(\overline{n}_{k})}{d \overline{n}_{k}} u_{k},
~~~\frac{d \overline{n}_{k}}{d \ln{g} } = 
-2 f(\overline{n}_{k}) \frac{ 1 - u_{k}^2}{1 + u_{k}^2}
\end{equation}
where $ \phi_{k}= \arctan{u_{k}}$. By trivial algebra 
we can get a differential relation between $u_k$ 
and $\overline{n}_{k}$ which can be 
exactly integrated with the result that 
$u_{k}^2 - f(\overline{n}_k) u_{k} + 1= 0$. By inverting 
this last relation we obtain two different solutions 
with equivalent physical properties, namely
\begin{equation}
u_{k}(\overline{n}_{k}) = \bigl[\frac{1}{2}(\sqrt{ 
\overline{n}_{k}(\overline{n}_{k}+ 1)} 
\pm \sqrt{\overline{n}_{k}(\overline{n}_{k} +1)  -4})\bigr].
\label{rel}
\end{equation}
If we choose the minus sign in Eq. (\ref{rel}) we obtain that 
$\phi_{k} \sim ( m + 1) \pi/2 $, $m= 0, 1,2...$
 with corrections of order 
$1/\overline{n}_k$ . In the opposite case 
 $\phi_{k} \sim \arctan{(\overline{n}_{k}/2)}$ within the same accuracy 
of the previous case(i.e. $1/\overline{n}_{k}$). 
By  using the relation between $u_{k}$ and $\overline{n}_{k}$ the condition
 $|(d\ln{g}/dx)(d f(\overline{n}_k)/d \overline{n}_k) 
\sin{2 \phi_{k}}| > 1$ is equivalent to $ x \laq 1$, 
if, as we are assuming, $|g'/g|$ vanishes 
as $\eta^{-2}$ for $\eta\rightarrow \pm \infty$ and it is, 
piece-wise, continuous.  
By inserting Eq. (\ref{rel}) into Eq. (\ref{IV})  a consistent solution 
can be obtained, in this case, if we integrate 
the system between $\eta_{f}$ and $\eta_{i}$ defined 
as the conformal times where 
$|(d\ln{g}/dx)(d f(\overline{n}_k)/d \overline{n}_k) 
\sin{2 \phi_{k}}|= 1$:
\begin{eqnarray}
&&\overline{n}_{k}(\eta_{f}) \sim 
\frac{1}{4} \biggl(\frac{g(\eta_{f})}{g(\eta_i)} 
- \frac{g(\eta_{i})}{g(\eta_{f})}\biggr)^2, 
\nonumber\\
&&\phi_{k}(\eta) \sim ( m + 1) \frac{\pi}{2}
 + {\cal O}(\frac{1}{\overline{n}_{k}(\eta)}),~~~m= 0, 1,2...
\label{xl1}
\end{eqnarray}
If $|(d\ln{g}/dx)(d f(\overline{n}_k)/d \overline{n}_k) 
\sin{2 \phi_{k}}| < 1$  (i. e. $x > 1$)  
the consistent solution of our system is given by
\begin{eqnarray}
&&\overline{n}_{k}(\eta_{f}) =
\sinh^2{\biggl(2 \int^{k\eta} \ln{g(x')} \sin{2 x'} dx'\biggr)}
\nonumber\\
&&\phi_{k} \sim - k\eta + \varphi_{k}, ~~~\varphi_{k} \simeq {\rm constant}.
\end{eqnarray}
If the coupling constant evolves 
continuously between $-\infty$ and $+\infty$ with a (global) maximum located 
at some time $\eta_r$ then, for $x> 1$,
 $\overline{n}_{k} \sim {\rm const.}$. Indeed 
by taking trial functions with bell-like shape for $|g'/g|$ we can show
that $\overline{n}_{k}$ oscillates around zero for large $\phi_{k}$.

From Eq.  (\ref{xl1}) 
the magnetic energy density of Eq. (\ref{endens}) 
can be computed in different scenarios and related 
to the ratio discussed in Eq. (\ref{dyn}). 
Recalling that the present frequency corresponding to 
$1$ Mpc is roughly $\omega_{\rm G} \sim 10^{-14}$ Hz, the 
ratio $r(\omega_{\rm G})$ can be estimated. Time evolution 
of gauge coupling during an inflationary phase 
produces rather large seeds $r(\omega_{\rm G}) \sim 10^{-12}$ \cite{dinng}.
Furthermore, pre-big bang models \cite{gabr} also lead to large seeds
which can be even  $r(\omega_{\rm G}) \sim 10^{-8}$ \cite{pbb1,pbb2}.
There is a difference in the models discussed in \cite{dinng} and
in \cite{pbb1,pbb2}. In the case of \cite{dinng} the gauge coupling is 
related to the expectation value of a scalar field which evolves during 
an inflationary phase of de Sitter (or quasi de Sitter type). This 
scalar degree of freedom is not the inflaton and it is 
not a source of the background geometry. On the contrary, in \cite{pbb1,pbb2}
the gauge coupling is a source of the evolution of the background geometry 
since it is connected to the expectation value of the dilaton field whose 
specific evolution dictates the nature of the pre-big bang solutions 
used in order to describe the dynamics of the Universe in its 
early stages.

\renewcommand{\theequation}{5.\arabic{equation}}
\setcounter{equation}{0} 
\section{EWPT and BAU}
In the previous Section it has been stressed that causal 
mechanisms have, in general problems with the correlation 
scale of the obtained field, while 
inflationary mechanisms may have problems with the seed
amplitude. In spite of this, it should be borne in mind 
that magnetic fields are generated over all physical scales compatible 
with the plasma dynamics at a given epoch. Hence, even if the
magnetic fields at large scales may be very 
minute, magnetic fields at smaller scales
may have a very interesting impact
on different moments of the life of the Universe. 
The physical picture we have in mind is then the 
following. Suppose that conformal 
invariance is broken at some stage 
in the evolution of the Universe, for instance 
thanks to the (effective) time variation 
of gauge couplings. Then, vacuum fluctuations 
will go outside the horizon and will be amplified. 
The amplified magnetic inhomogeneities  will 
re-enter (crossing the horizon a second time) 
during different moments of the life of the Universe
and, in particular, 
even before the BBN epoch. 

In the following various effects of these 
magnetic fields will be considered starting with 
the EW epoch. The electroweak epoch 
occurs when the temperature of the plasma was roughly 
$T\sim T_{c} \sim 100$ GeV. The physical size 
of the horizon was, at that time , $H_{\rm ew}^{-1} 
\sim 3 $ cm. The electroweak epoch occurs, 
approximately, when the Universe was $10^{-11}$ sec old.

At small temperatures and small densities of different 
fermionic charges the $SU_{L}(2) \otimes U_{Y}(1)$  is 
broken down to the $U_{\rm em}(1)$ and the long range fields which can 
survive in the plasma are the ordinary magnetic fields. 
However, for sufficiently high temperatures the 
$SU_{L}(2) \otimes U_{Y}(1)$ is restored and
non-screened vector modes correspond to hypermagnetic fields. 
At the electroweak epoch the typical size of the horizon is of the 
order of $3$ cm . The typical diffusion scale is of the order of $10^{-9}$ cm.
Therefore, over roughly eight orders of magnitude hypermagnetic fields can 
be present in the plasma without being dissipated \cite{mg3}. 
The evolution of hypermagnetic fields can be obtained from the  anomalous 
magnetohydrodynamical (AMHD) 
equations. The AMHD equations generalize the treatment 
of plasma effects involving hypermagnetic fields to the case 
of finite fermionic density\cite{mg4}. 

Depending on their topology, hypermagnetic fields can have various 
consequences
\cite{mg3,mg4}. 
If the hypermagnetic flux lines have a trivial topology they can have an impact on 
the phase diagram of the electroweak phase transition \cite{h,h2}. 
If the topology of hypermagnetic fields is non trivial, hypermagnetic knots 
can be formed \cite{kn1} and, under specific conditions, the BAU
can be generated \cite{kn2}. The gauge field fluctuations 
produced as a result of the parametric amplification 
of vacuum fluctuations always lead to hypermagnetic fields with 
topologically trivial structure (i.e. with zero magnetic helicity 
and gyrotropy). However, 
thanks to pseudo-scalar couplings, a topologically trivial 
background of hypermagnetic flux lines may lead to a 
non-zero magnetic gyrotropy and, hence, 
to some kind of hypermagnetic knot with topologically 
non-trivial structure.

A classical hypermagnetic background in the symmetric phase of the EW theory can produce
interesting amounts of gravitational radiation  
 in a frequency range between $10^{-4}$ Hz
and the kHz. The lower tail falls into the LISA window while the 
higher tail falls in the VIRGO/LIGO window. For the hypermagnetic background 
required in order to seed the BAU the amplitude of the obtained GW 
can be even six orders of magnitude larger than the 
inflationary predictions. In this context, the mechanism of 
baryon asymmetry generation is connected with GW production \cite{kn1,kn2}.

\subsection{Hypermagnetic knots}
It is  possible to construct hypermagnetic knot configurations 
with finite energy and helicity which are localized in space and within 
typical distance scale  $L_{s}$. 
Let us consider in fact the following configuration
in spherical coordinates \cite{kn2}
\begin{eqnarray}
{\cal Y}_{r}({\cal R},\theta) &=& - \frac{2 B_0}{ \pi L_{s}}
\frac{\cos{\theta} }{\bigl[{\cal R}^2 +1\bigr]^2},
\nonumber\\
{\cal Y}_{\theta}({\cal R},\theta) &=& \frac{2 B_0}
{ \pi L_{s}} \frac{ \sin{\theta}}{\bigl[ {\cal
R}^2 + 1\bigr]^2},
\nonumber\\
{\cal Y}_{\phi}({\cal R},\theta) &=& - \frac{ 2 B_0}{ \pi L_{s}} \frac{ n
{\cal
R}\sin{\theta}}{\bigl[{\cal R}^2 + 1\bigr]^2},
\label{conf2}
\end{eqnarray}
where ${\cal R}= r/L_{s}$ is the rescaled radius and $B_{0}$ is some 
dimensionless amplitude and $n$ is just an integer number 
whose physical interpretation will become clear in a moment. 
The hypermagnetic field can be easily computed 
from the previous expression and it is 
\begin{eqnarray}
&&{\cal H}_{r}({\cal R},\theta) = - \frac{4 B_{0}}{\pi~ L_{s}^2}\frac{n
\cos{\theta}}{\bigl[  {\cal
R}^2 + 1\bigr]^2},
\nonumber\\
&&{\cal H}_{\theta}({\cal R}, \theta) = - \frac{4 B_{0}}{\pi~
L_{s}^2}\frac{{\cal R}^2 -1}{\bigl[
{\cal R}^2 + 1\bigr]^3}n \sin{\theta},
\nonumber\\
&&{\cal H}_{\phi}({\cal R}, \theta) = 
- \frac{8 B_0}{ \pi~ L_{s}^2}\frac{ {\cal R}
\sin{\theta}}{\bigl[
{\cal R}^2 + 1\bigr]^3}.
\label{knot}
\end{eqnarray}
The poloidal and toroidal components of $\vec{{\cal H}}$ can be usefully 
expressed as $\vec{{\cal H}}_{p} = 
{\cal H}_{r} \vec{e}_{r} + {\cal H}_{\theta} \vec{e}_{\theta} $ 
and $\vec{\cal H}_{t}= {\cal H}_{\phi} \vec{e}_{\phi}$.
The Chern-Simons number is finite and it is given by 
\begin{equation}
N_{CS} =\frac{g'^2}{32\pi^2}
\int_{V} \vec{{\cal Y}} \cdot \vec{{\cal H}}_{{\cal Y}} d^3 x=
\frac{g'^2}{32\pi^2} \int_{0}^{\infty}
\frac{ 8 n
B^2_0}{\pi^2} \frac{ {\cal R}^2 d {\cal R}}{\bigr[ {\cal R}^2 +
1\bigl]^4} = \frac{g'^2 n B^2_0}{32 \pi^2}.
\label{CS}
\end{equation}
We can also compute the total helicity of the configuration namely
\begin{equation}
\int_{V} \vec{{\cal H}}_{Y}
 \cdot \vec{\nabla} \times \vec{{\cal H}}_{Y} d^3 x=
\frac{256~B^2_0~n}{\pi L^2 } \int_{0}^{\infty} \frac{ {\cal R}^2 d
{\cal R}}{(1 + {\cal R}^2)^5} = \frac{5 B^2_0 n}{L_s^2}.
\label{helic}
\end{equation}
We can compute also the total energy of the field
\begin{equation}
E = \frac{1}{2}\int_{V} d^3 x |\vec{{\cal H}}_{Y}|^2 = \frac{B^2_0}{2~L_{s}}
(n^2 + 1).
\end{equation}
and we discover that it is proportional to $n^2$.
 This means that one way of increasing the total energy of
 the field is to increase the number of knots and twists in the flux lines.
We can also have some real space pictures of the core of the knot
 (i.e. ${\cal R} = r/L_{s}<1$).
This type of configurations can be also obtained by projecting a 
non-Abelian SU(2) (vacuum) gauge field on a fixed electromagnetic 
direction \cite{JPI} \footnote{ In order to interpret these solutions it is 
very interesting to make use of the Clebsh decomposition. The 
implications of this  decomposition (beyond the hydrodynamical context, where 
it was originally discovered) have been recently discussed (see \cite{JPI2} 
and references therein). I thank R. Jackiw for interesting discussions about 
this point.}
These configurations have been also 
studied in \cite{ad1,ad2}.  In particular, in \cite{ad2}, the relaxation 
of HK has been investigated with a technique different from the one employed
 in \cite{kn1,kn2} but with similar results. The problem of scattering of 
fermions in the background of hypermagnetic fields has been also studied in 
\cite{ay}.

Topologically non-trivial 
configurations of the hypermagnetic flux lines lead to the formation 
of hypermagnetic knots (HK) whose decay 
might seed the Baryon Asymmetry of the Universe (BAU).
HK can be dynamically generated  provided
a topologically trivial (i.e. stochastic)
distribution of flux lines is already present 
in the symmetric phase of the electroweak (EW) theory \cite{kn1,kn2}. 
In spite of
the mechanism generating the HK, their typical size
must exceed the diffusivity length scale. 
In the  minimal standard model (MSM) (but not necessarily 
 in its supersymmetric extension) HK are washed out. 

The importance of the topological properties 
of long range (Abelian) hypercharge magnetic fields has been  
stressed in the  past \cite{vi,ru,ru2,ru3}.
In \cite{m1} it was argued that if the spectrum of hypermagnetic 
fields is dominated by parity non-invariant Chern-Simons 
(CS) condensates, the BAU could be the result of their decay. 
Most of the mechanisms
often invoked for the origin of large scale magnetic fields in the early 
Universe seem to imply the production of topologically trivial (i.e. 
stochastic) configurations of magnetic fields \cite{s1,s2,s3}.

\subsection{Hypermagnetic knots and BAU}

Suppose that the EW plasma is filled, for $T> T_{c}$ 
with topologically trivial hypermagnetic fields $\vec{\cal H}_{Y}$, 
which  can be physically pictured as a 
collection of flux tubes (closed because of the transversality 
of the field  lines)  evolving independently without 
breaking  or intersecting with each other. If the field
 distribution is topologically 
trivial (i.e. $\langle\vec{\cal H}_{Y} \cdot\vec{\nabla} 
\times\vec{\cal H}_{Y}\rangle =0$) parity is  a  good symmetry 
of the plasma and the field can be completely homogeneous. 
We name hypermagnetic knots those CS condensates carrying 
a non vanishing (averaged)  hypermagnetic helicity
(i.e.  $\langle\vec{\cal H}_{Y} \cdot\vec{\nabla} 
\times\vec{\cal H}_{Y}\rangle \neq 0$). 
If $\langle\vec{\cal H}_{Y} \cdot\vec{\nabla} 
\times\vec{\cal H}_{Y}\rangle \neq 0$  parity is  broken for scales 
comparable with the size of the HK,
 the flux lines are knotted and the field $\vec{{\cal H}}_{Y}$ 
cannot be completely homogeneous.  

In order to seed the BAU a network of HK should be present at high
temperatures \cite{mg3,mg4,m1}. In fact
for temperatures larger than $T_{c}$
 the fermionic number is stored both in HK 
and in real fermions.  For $T<T_{c}$, 
the HK should release real fermions 
since the ordinary magnetic fields (present {\em after} EW 
symmetry breaking) do not carry fermionic number.
If the EWPT is strongly first order the decay of the HK 
can offer some seeds for the BAU generation \cite{m1}.
This last condition can be met in the 
minimal supersymmetric standard model (MSSM) \cite{PT,mssm,mssm1,mssm2}.

Under these hypotheses the integration of the $U(1)_{Y}$ 
anomaly equation \cite{m1}
gives the CS number density carried by the HK
which is in turn related to the density of baryonic number $n_{B}$
for the case of $n_{f}$ fermionic generations.
\begin{equation}
\frac{n_{B}}{s}(t_{c})=
\frac{\alpha'}{2\pi\sigma_c}\frac{n_f}{s}
\frac{\langle{\vec{{\cal H}}}_{Y}\cdot \vec{\nabla}\times
{\vec{{\cal H}}}_{Y}\rangle}{\Gamma + \Gamma_{{\cal H}}}
\frac{M_{0}\Gamma}{T^2_c},~~\alpha' = \frac{g'^2}{4\pi}
\label{BAU}
\end{equation}
($g'$ is the $U(1)_{Y}$ coupling and $s = (2/45) \pi^2 N_{eff}T^3$ 
is the entropy density; $N_{eff}$, at $T_{c}$,
 is $106.75$ in the MSM;
$M_{0}= M_{P}/1.66 \sqrt{N_eff} \simeq 7.1 \times 10^{17} {\rm GeV}$).
In Eq. (\ref{BAU}) $\Gamma$ is the perturbative rate of the 
right electron chirality 
flip processes  (i.e. 
scattering of right electrons with the Higgs and gauge bosons and with 
the top quarks because of their large Yukawa coupling) which 
are the slowest reactions in the plasma and 
\begin{equation}
\Gamma_{{\cal H}} = \frac{783}{22} \frac{\alpha'^2}{\sigma_{c} \pi^2} 
\frac{|\vec{{\cal H}}_{Y}|^2}{T_{c}^2}
\end{equation}
is the rate of right electron dilution induced by the presence of a
 hypermagnetic field. In the MSM we have that 
$\Gamma < \Gamma_{{\cal H}}$ \cite{fl} 
whereas in the MSSM $\Gamma$ can naturally 
be larger than $\Gamma_{{\cal H}}$. 
Unfortunately, in the MSM 
a hypermagnetic field can modify the phase diagram of the phase transition 
but cannot make the phase transition strongly first order for large masses of
the Higgs boson \cite {h}. 
Therefore, we will concentrate on the case $\Gamma > \Gamma_{\cal H}$ and we
 will show that in the opposite limit the BAU will be anyway small 
even if some (presently unknown) mechanism would make the EWPT strongly 
first order in the MSM.

HK can be dynamically generated \cite{kn1,kn2} (see also \cite{fr}).
Gauge-invariance 
and transversality of the magnetic fields suggest
 that perhaps
 the only way of producing  $\langle\vec{\cal H}_{Y} \cdot\vec{\nabla} 
\times\vec{\cal H}_{Y}\rangle \neq 0$ is to 
postulate, a time-dependent interaction between the two (physical) 
 polarizations of the hypercharge field $Y_{\alpha}$.
Having defined the Abelian field strength  $Y_{\alpha\beta} = 
\nabla_{[\alpha} Y_{\beta]}$ and its dual $\tilde{Y}_{\alpha\beta}$ such an 
 interaction can be described, in curved space, by the 
Lagrangian density
\begin{equation}
L_{eff}= \sqrt{-g} \biggl[ 
-\frac{1}{4}Y_{\alpha\beta} Y^{\alpha\beta} + 
c\frac{\psi}{4 M}
Y_{\alpha\beta}\tilde{Y}^{\alpha\beta}\biggr].
\label{action}
\end{equation}
where $g_{\mu\nu}$ is the metric tensor and $g$ its determinant, 
$c$ is the coupling  constant and $M$ is a typical scale.
This  interaction 
is plausible  if the $U(1)_{Y}$ anomaly is coupled, 
(in the symmetric phase of the EW theory ) to  dynamical 
pseudoscalar particles $\psi$ (like the axial Higgs of the MSSM).
Thanks to the presence of pseudoscalar particles, 
the two polarizations of $\vec{{\cal H}}_{Y}$ 
evolve in a slightly different way
producing, ultimately, inhomogeneous HK.

Suppose that 
an inflationary phase with $a(\tau) \sim \tau^{-1}$  is continuously 
matched, at the transition time $\tau_1$, to a radiation dominated 
phase where $a(\tau)\sim \tau$. Consider then a massive pseudoscalar 
field $\psi$ which oscillates during the last  stages 
of the inflationary evolution
with typical amplitude $\psi_0 \sim M$.
As a result of the inflationary evolution
 $|\vec{\nabla}\psi| \ll \psi'$. Consequently, the 
phase of $\psi$  can  get frozen. 
Provided  the 
pseudoscalar mass $m$ is larger than the inflationary curvature  scale 
$H_{i}\sim {\rm const.}$, the $\psi$ oscillations are converted, 
at the end of the
 quasi-de Sitter stage, in a net helicity arising as a result 
of the different evolution of the two (circularly polarized) vector potentials
\begin{eqnarray}
&& {Y}_{\pm}'' + \sigma Y'_{\pm}+
\omega_{\pm}^2 {Y}_{\pm} =0,~~\vec{H}_{Y} = \vec{\nabla} 
\times \vec{Y}
\label{Y}\\
&& \omega_{\pm}^2 = k^2 \mp k \frac{c}{M} a \dot{\psi}
\end{eqnarray}
(where we denoted with $\vec{H}_{Y} = a^2 {\vec{\cal H}}_{Y}$  the 
curved space fields and with $\sigma= \sigma_c a$ the rescaled 
hyperconductivity; the prime denotes derivation with respect to 
conformal time $\tau$ whereas the over-dot denotes differentiation 
with respect to cosmic time $t$). 

Since $\omega_{+}\neq \omega_{-}$ the helicity 
gets amplified according to Eq. (\ref{Y}). 
There are two important points to stress in this context. 
First of all the plasma effects as well as the finite density effects 
are important. This means, in practical terms, that the dissipation scales 
of the problem should be borne in mind. This has not always 
been done. The second point is related to the first. By treating, consistently, 
the plasma and finite density effects in the context of AMHD \cite{mg3,kn1,kn2}, 
one realizes that the pseudoscalar coupling of $\psi$ together 
with the coupling of the chemical potential are not 
sufficient in order to seed the BAU {\em unless} some 
hypermagnetic background is already present. 
In other words the scenario which leads to the 
generation of the BAU is the following. 
Correlation in the hypermagnetic fields are generated 
outside the horizon during inflation by direct breaking 
of conformal invariance. An example in this direction 
has been given in the 
previous Section. Then hypermagnetic fields 
re-entering at the electroweak epoch will participate 
in the dynamics and, in particular, they will 
feel the effect of the anomalous coupling either to $\psi$ or to 
the chemical potential. The effect of the anomalous coupling 
will not be to amplify the hypermagnetic background. The 
anomalous coupling will only make the topology of the hypermagnetic 
flux lines non-trivial. So the statement is that 
{\em if} conformal invariance is broken and, {\em if} hypermagnetic 
have anomalous couplings, {\em then} a BAU $\gaq 10^{-10}$ can be achieved
without spoiling the standard cosmological evolution \cite{kn1,kn2}.
It is worth mentioning that this type of scenario may be motivated by 
the low energy string effective action where, by supersymmetry, 
Kalb-Ramond axions and dilatons are coupled, respectively, to 
$Y_{\mu\nu}\tilde{Y}^{\mu\nu}$ and to the gauge kinetic term \cite{kr}.
It is interesting to notice that, in this scenario, the value 
of the BAU is determined by various particle physics parameters but also 
by the ratio of the hypermagnetic energy density over the 
energy density sitting in radiation during 
the electroweak epoch, namely, using the language 
of the previous Sections \cite{kn1,kn2}, 
\begin{equation}
\frac{n_{B}}{s} \propto  r.
\end{equation}
In order to get a sizable BAU $r$ should be at least $10^{-3}$ 
if the anomalous coupling operates during a radiation 
phase. The value of $r$ could be  smaller in models where 
the anomalous coupling is relevant during a low scale inflationary phase \cite{kn2}.

\renewcommand{\theequation}{6.\arabic{equation}}
\setcounter{equation}{0} 
\section{BBN and matter--antimatter fluctuations}
Large scale magnetic fields possibly present at the BBN epoch 
can have an impact on the light nuclei formation. By reversing 
the argument, the success of BBN can be used in order 
to bound the magnetic energy density possibly present at 
the time of formation of light nuclei.

These bounds are qualitatively  different from the ones previously quoted and
coming, alternatively, from homogeneity \cite{hom} and isotropy 
\cite{iso} of the background geometry at the BBN time.
As elaborated in slightly different frameworks
 through the years \cite{bbn1,bbn2,bbn3,bbn4,bbn5}, magnetic fields 
possibly present at the BBN epoch could have a twofold effect. 
On one hand they could 
enhance the rate of reactions (with an effect
proportional to $\alpha \rho_{B}$) 
and, on the other hand they could 
artificially increase the expansion rate (with an effect 
proportional to $\rho_{B}$). It turns out that 
the latter effect is probably the most relevant \cite{bbn4}. 
 In order to prevent the Universe 
from expanding too fast at the BBN epoch 
$\rho_{B} < 0.27 \rho_{\nu}$  where  $\rho_{\nu}$ is the 
energy density contributed by the standard three neutrinos for $T< 1$ MeV.

In the previous Section the case of a 
{\em topologically non-trivial} 
hypermagnetic background has been considered (i.e. 
$\langle \vec{{\cal H}}_{Y} \cdot \vec{\nabla} 
\times \vec{{\cal H}}_{Y} \rangle \neq 0$).  
In this Section we will instead assume that the 
hypermagnetic background is {\em topologically 
trivial} (i.e. $\langle \vec{{\cal H}}_{Y} \cdot 
\vec{\nabla} 
\times \vec{{\cal H}}_{Y} \rangle = 0$). In this 
case, fluctuations in the baryon to entropy ratio
will be induced since 
$\langle(\vec{{\cal H}}_{Y} \cdot \vec{\nabla} 
\times \vec{{\cal H}}_{Y})^2 \rangle \neq 0$.
These fluctuations are of {\em isocurvature}
type and can be  related to the spectrum 
of hypermagnetic fields at the EWPT. 
Defining as 
\begin{equation}
\Delta(r,t_{c}) = \sqrt{\langle\delta
\left(\frac{n_{B}}{s}\right)(\vec{x}, t_{c})
\delta\left(\frac{n_{B}}{s}\right)(\vec{x}+\vec{r},
t_{c})\rangle}, 
\label{13}
\end{equation}
the fluctuations in the baryon to entropy ratio 
at $t={t_c}$ \cite{mg4}, the value of $\Delta(r,t_{c})$
can be ralated to the hypermagnetic spectrum
which is determined in terms of its amplitude 
$\xi$ and its slope $\epsilon$.
A physically realistic situation corresponds to the case in
which the Green's functions of the magnetic hypercharge fields decay at
large distance (i. e. $\epsilon> 0$) and this would
imply either ``blue''( $\epsilon \geq 0$ ) or ``violet''
($\epsilon \gg 1$) energy spectra. The case of ``red'' spectra
($\epsilon < 0$) will then be left out of our discussion. The flat
spectrum corresponds to $\epsilon \ll 1$ and may appear quite naturally in
string cosmological models \cite{pbb1,pbb2}.

\begin{figure}[htb]
    \centering
    \includegraphics[height=2.5in]{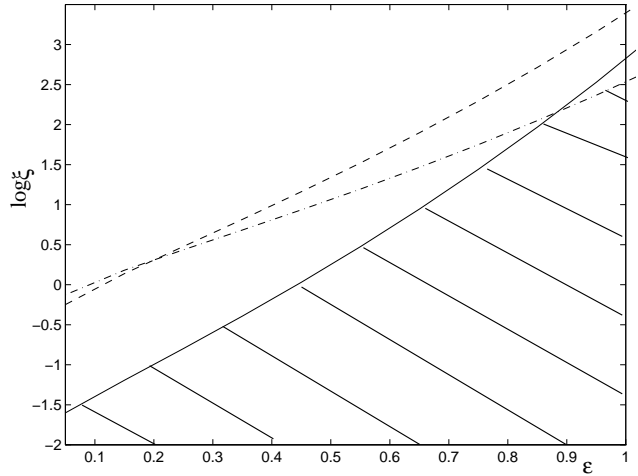}
    \caption{The parameter space of the hypermagnetic 
background in the case 
$\Delta(r, t_c) < n_{B}/s$ for $r> r_{\rm n}$ (full line).}
    \label{F2}
\end{figure}

Since the fluctuations in the baryon to entropy
ratio are not positive definite, they will induce 
fluctuations in the baryon to photon ratio, $\eta$
 at the BBN epoch. The possible effect of 
matter--antimatter fluctuations on BBN 
depends on the typical scale of of the 
baryon to entropy ratio at the electroweak epoch.
Recalling that for $T\sim T_{c}\sim 100 $ GeV 
the size of the electrowek horizon is 
approximately $3 $ cm, fluctuations whose 
scale is well inside the EW horizon at 
$T_{c}$ have dissipated by the BBN 
time through (anti)neutron diffusion.
The neutron diffusion scale at $T_{c}$ is 
\begin{equation}
 r_{\rm n}(T_{c})  =0.3 ~~{\rm cm}.
\end{equation}
 The neutron diffusion scale
at $T=1 ~{\rm keV}$ is  $10^{5}$ m, while, today 
it is $10^{-5}$ pc, i.e. of the order of the astronomical unit.
Matter--antimatter fluctuations smaller than $10^{5}$ m annihilate 
before neutrino decoupling and have no effect on BBN.
\begin{figure}
\begin{center}
\begin{tabular}{|c|c|}
      \hline
      \hbox{\epsfxsize = 7.5 cm  \epsffile{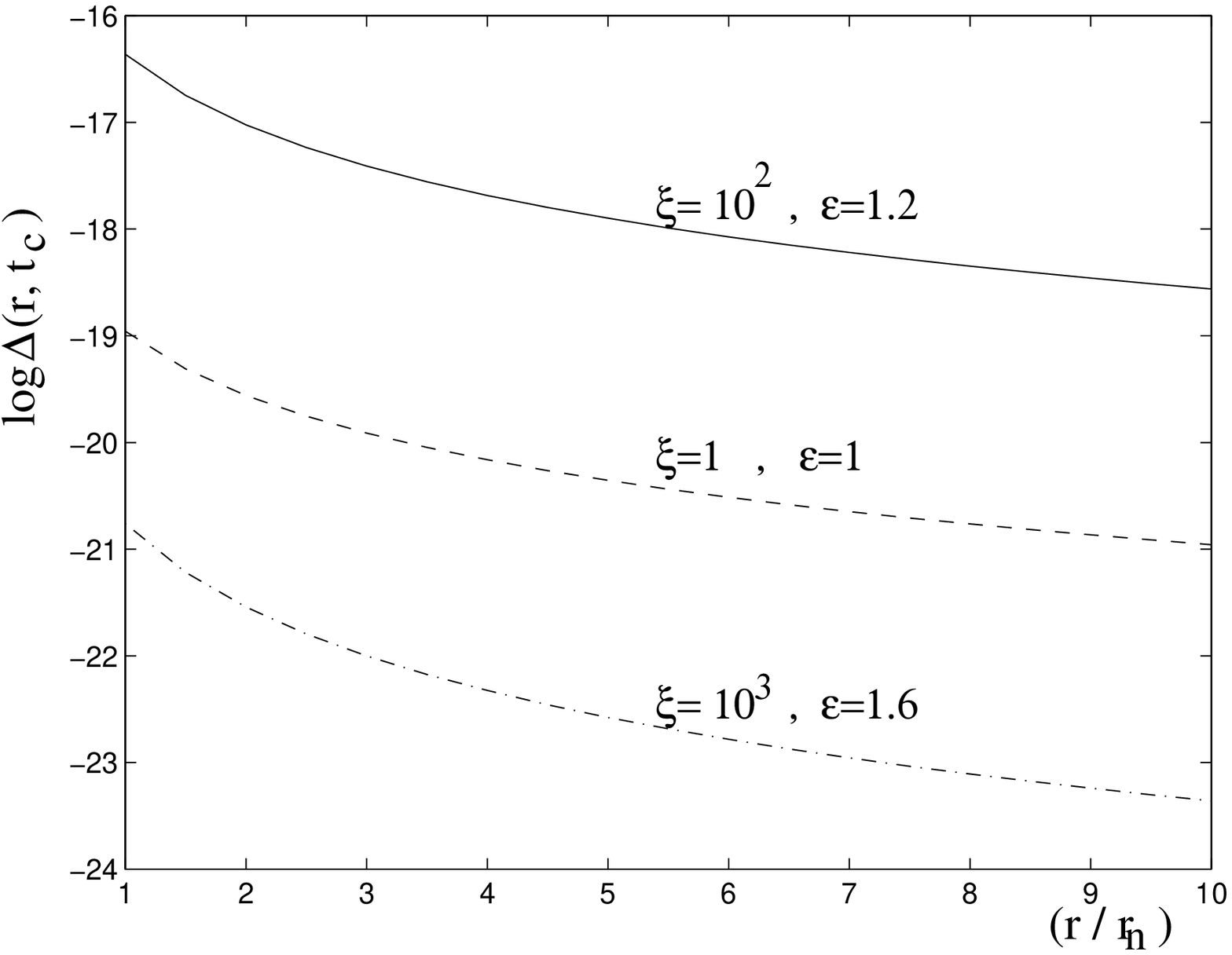}} &
      \hbox{\epsfxsize = 7.5 cm  \epsffile{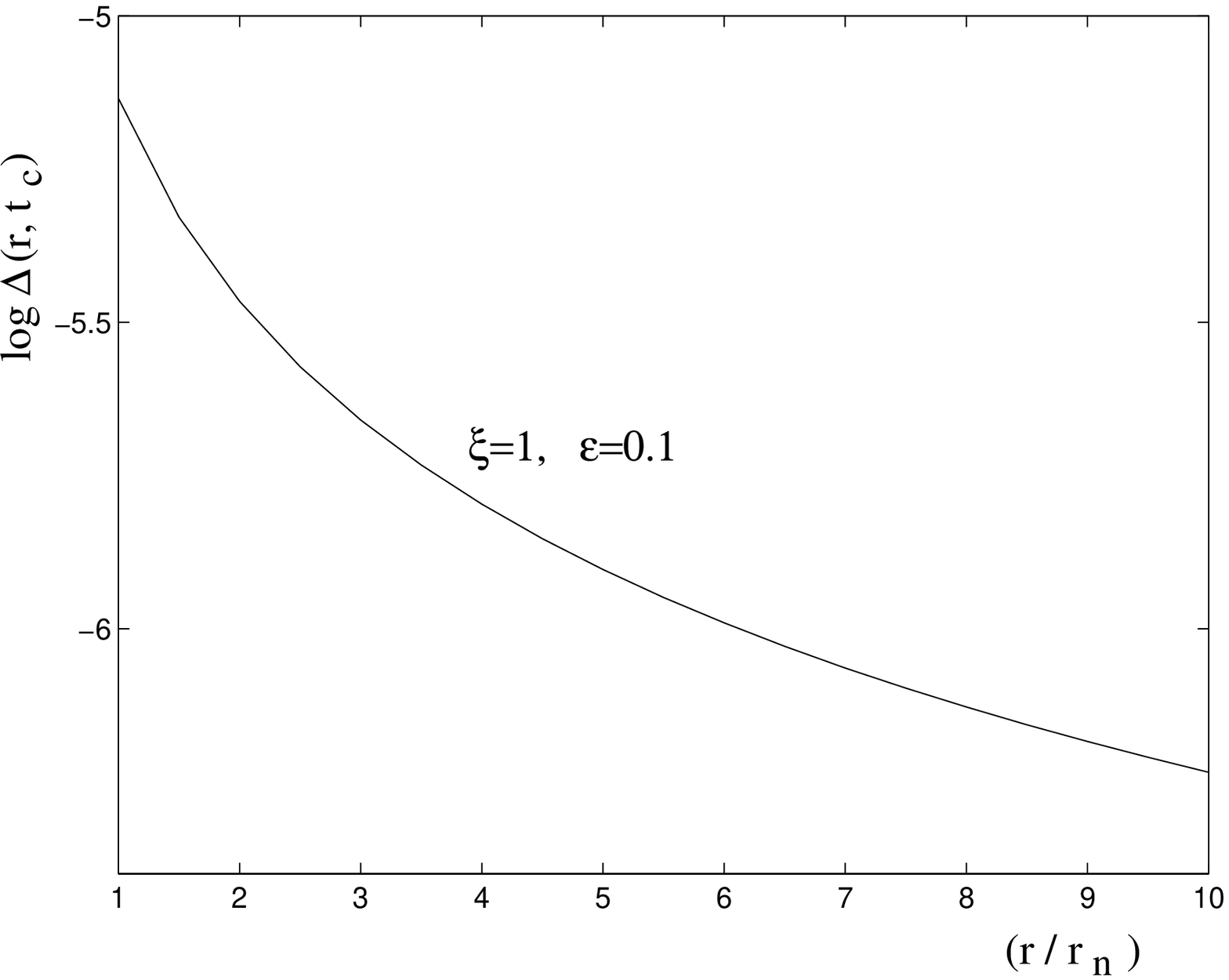}}\\
      \hline
\end{tabular}
\end{center}
\caption{We report the value of the baryon number 
fluctuations for different parameters of the hypermagnetic
background $\xi$ and $\epsilon$. }
\end{figure}
Two possibilities can then be envisaged. 
We could require that the matter--antimatter 
fluctuations (for scales $r\geq r_{\rm n}$) 
are small. This will then imply a 
bound, in the ($\xi$,$\epsilon$) plane on the strength 
of the hypermagnetic background. In Fig. \ref{F2} 
such an exclusion plot is reported with the full
line. With the dashed line the bound implied by 
the increase in the expansion rate is reported. 
Finally with the dot-dashed line the 
critical density bound is illustrated 
for the same hypermagnetic background.
The second possibility is to study the  effects 
of large matter--matter domains. 
These studies led to a slightly different 
scenario of BBN \cite{mam1}, namely BBN 
with matter--antimatter regions
\cite{mam2,mam3,mam4,mam4a,mam4b,mam5}. 
The idea is to discuss BBN in the presence 
of spherically symmetric regions 
of anti--matter characterized by their 
radius $r_{\rm A}$ and by the parameter $R$, i.e. 
the matter/antimatter ratio.
Furthermore, in this scenario the net baryon-to-photon
ratio,  $\eta$, is positive definite and non zero. 
Antimatter domains larger than $10^{5}$ m at $1$ keV may survive 
until BBN and their dissipation has been 
analyzed in detail in \cite{mam2,mam3,mam4,mam4a,mam4b,mam5}.
Antimatter domains  in the range 
\begin{equation}
10^{5} ~~{\rm m}~ \laq 
r_{\rm A} ~\laq~ 10^{7} ~~{\rm m}
\label{range}
\end{equation}
 at 1 keV annihilate 
before BBN for temperatures between 70 keV and $1$ MeV.
Since the antineutrons annihilate on neutrons, the neutron 
to proton ratio gets smaller. As a consequence, the 
$^{4}{\rm He}$ abundance gets reduced if compared to the 
standard BBN scenario. The maximal scale of matter--antimatter
fluctuations is determined by the constraints following 
from possible distortions of the CMB spectrum.
The largest scale is of the order of $100$ pc (today), corresponding 
to $10^{12}$ m at $1$ keV.  
\begin{figure}[htb]
    \centering
    \includegraphics[height=2.5in]{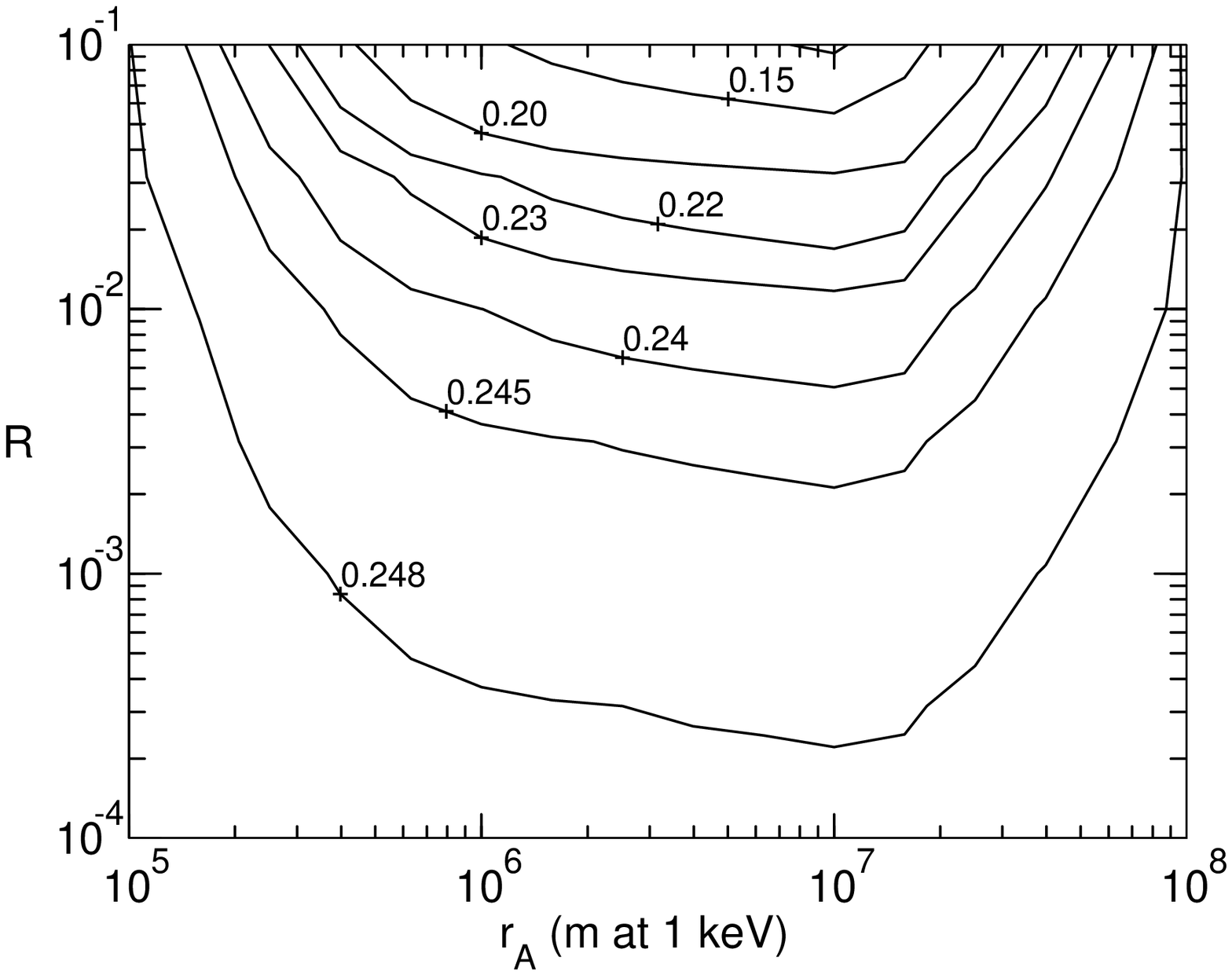}
    \caption{From \cite{mam6} the $^{4}{\rm He}$ yield is illustrated 
in the $(R,r_{\rm A})$ plane for $\eta = 6\times 10^{-10}$. As the 
matter/antimatter ratio decreases, we recover the 
standard $^{4}{\rm He}$ yield.}
    \label{F3}
\end{figure}
Suppose that matter--antimatter regions are present 
in the range of Eq. (\ref{range}). Then the abundance 
of $^{4}{\rm He}$ get reduced. The yield of $^{4} {\rm He}$ are 
reported as a function of $R$, the matter--antimatter ratio
and $r_{\rm A}$. Now, we do know that by adding 
extra-relativistic species the $^{4} {\rm He}$ can be 
increased since the Universe expansion gets larger. 
Then the conclusion is that BBN with 
matter--antimatter domains allows for a larger number 
of extra-relativistic species if compared to the 
standard BBN scenario. This observation 
may have implications for the upper bounds on the stochastic 
GW backgrounds of cosmological origin \cite{mam6} since 
the extra-relativistic species present at the BBN
epoch can indeed be interpreted as relic gravitons.

\renewcommand{\theequation}{7.\arabic{equation}}
\setcounter{equation}{0} 
\section{GW backgrounds}

If a hypermagnetic background is present for $T> T_c$, then, as also discussed
in \cite{mmm} in the context of ordinary MHD,  the energy momentum tensor 
will acquire a small anisotropic component which will source the evolution 
equation of the tensor fluctuations $h_{\mu\nu}$ of the metric $g_{\mu\nu}$: 
\begin{equation}
h_{ij}'' + 2 {\cal H} h_{ij}' - \nabla^2 h_{ij} = - 16 \pi G
\tau^{(T)}_{ij}.
\label{GWeq}
\end{equation}
where $\tau^{(T)}_{ij}$ is the {\em tensor} component of the 
{\em energy-momentum tensor} \cite{mmm} 
of the hypermagnetic fields. Suppose now, as assumed in \cite{h} that 
$|\vec{{\cal H}}|$ has constant amplitude and that it is also 
homogeneous. Then 
as argued in  \cite{griru} we can easily deduce 
the critical fraction of energy density  present today in relic gravitons 
of EW origin 
\begin{equation}
\Omega_{\rm gw}(t_0) = \frac{\rho_{\rm gw}}{\rho_c} 
\simeq z^{-1}_{{\rm eq}}
r^2,~~\rho_{c}(T_{c})\simeq N_{\rm eff} T^4_{c }
\end{equation}
($z_{\rm eq}=6000$ is the redshift from the time of matter-radiation,
 equality to the present time 
$t_0$). Because of the structure of the AMHD equations, stable 
hypermagnetic fields will be present not only for 
$\omega_{\rm ew}\sim k_{\rm ew}/a$ but 
for all the range $\omega_{{\rm ew}} <\omega< \omega_{\sigma}$ where 
$\omega_{\sigma}$ is the diffusivity frequency. Let us assume, 
 for instance, that $T_{c} \sim 100 $ GeV and $N_{eff} = 106.75$. 
Then, the (present) values of 
$\omega_{\rm ew}$ is 
\begin{equation}
\omega_{\rm ew } (t_0) \simeq 2.01 \times 10^{-7} \biggl( \frac{T_{c}}{1 {\rm GeV}} \biggr) 
\biggl(\frac{ N_{\rm eff}}{100} \biggr)^{1/6} {\rm Hz} .
\end{equation}
Thus, $\omega_{\sigma}(t_0) \sim 10^{8} \omega_{\rm ew} $. Suppose now that 
$T_{c} \sim 100$ GeV; than we will have that $\omega_{\rm ew}(t_0) \sim 10^{-5}$ Hz. 
Suppose now, as assumed in \cite{h}, that 
\begin{equation}
|\vec{{\cal H}}|/T_{c}^2 \gaq 0.3.
\end{equation} 
This requirement imposes $ r \simeq 0.1$--$0.001$ and, consequently, 
\begin{equation}
h_0^2 \Omega_{\rm GW} \simeq 10^{-7} - 10^{-8}.
\end{equation}
Notice that this signal would occurr in a (present) frequency 
range between $10^{-5}$ and $10^{3}$ Hz. This signal 
satisfies the presently available phenomenological 
bounds on the graviton backgrounds of primordial origin.
The pulsar timing bound ( which applies for present 
frequencies $\omega_{P} \sim 10^{-8}$ Hz and implies 
$h_0^2 \Omega_{\rm GW} \leq 10^{-8}$) is automatically satisfied
since our hypermagnetic background is defined for $10^{-5} {\rm Hz} 
\leq \omega \leq 10^{3} {\rm Hz}$. The large scale bounds would imply 
$h_0^2 \Omega_{\rm GW} < 7 \times 10^{-11}$ but a at much lower frequency 
(i.e. $10^{-18 }$ Hz). The signal discussed here is completely 
absent for frequencies $\omega < \omega_{\rm ew}$. Notice that 
this signal is clearly distinguishable from other stochastic 
backgrounds occurring at much higher frequencies (GHz region) 
like the ones predicted by quintessential inflation \cite{gw1}.
It is equally distinguishable from signals due to 
pre-big-bang cosmology (mainly in the window of
ground based interferometers \cite{gw2}).
The frequency of operation of the interferometric devices 
(VIRGO/LIGO) is located between few Hz and 10 kHz \cite{gw2}.
 The frequency of operation 
of LISA is well below the Hz (i.e. $10^{-3} $Hz, approximately). 
 In this model the signal 
can be located both in the LISA window and in the VIRGO/LIGO window
due to the hierarchy between the hypermagnetic diffusivity scale and the 
horizon scale at the phase transition \cite{kn1,kn2}.
\begin{figure}[htb]
    \centering
    \includegraphics[height=2.5in]{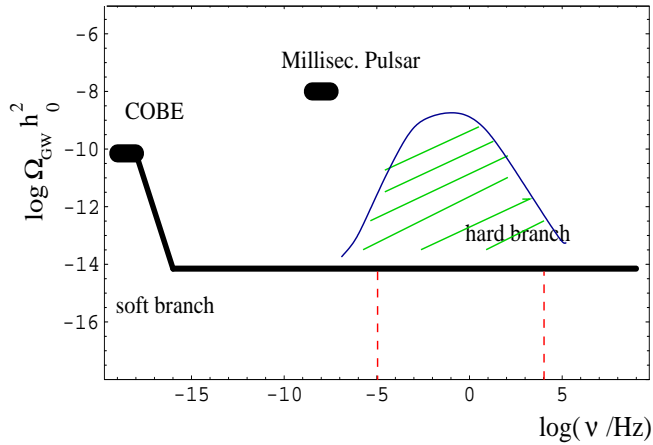}
    \caption{The stochastic background of GW produced by 
inflationary models with flat logarithmic energy spectrum, 
illustrated together with the GW background of hypermagnetic origin. The 
frequencies marked with dashed lines correspond to the electroweak frequency and to the 
hypermagnetic diffusivity frequency.}
    \label{F4}
\end{figure}
\renewcommand{\theequation}{8.\arabic{equation}}
\setcounter{equation}{0} 
\section{Faraday rotation of CMB ?}

Large scale magnetic fields present at the decoupling epoch can have 
various consequences. For instance they can induce fluctuations in the 
CMB \cite{pbb1,pbb2}, they can distort the Planckian spectrum of CMB 
\cite{jed}, 
they can distort the acoustic peaks of  CMB anisotropies \cite{rg} and 
they can also depolarize CMB \cite{FR}.

The polarization of the CMB
represents a very interesting observable which has been extensively
investigated in the past both from the theoretical \cite{1a} and
experimental points of view \cite{2a}. Forthcoming satellite missions
like PLANCK \cite{3a} seem to be able to achieve a level
of sensitivity which will enrich decisively our experimental knowledge
of the CMB polarization with new direct measurements. 

If the background geometry of the universe is homogeneous but not
isotropic the CMB is naturally polarized \cite{1a}. 
This phenomenon occurs, for example, in  Bianchi-type I models \cite{4a}.
On the other hand if the background geometry is homogeneous and
isotropic (like in the Friedmann-Robertson-Walker  case) it seems very
reasonable that the CMB acquires a small degree of linear
polarization provided the radiation field has a non-vanishing
quadrupole component at the moment of last scattering \cite{5a}.

Before decoupling photons, baryons and electrons form a unique fluid
which possesses only monopole and dipole moments, but not
quadrupole. Needless to say, in a homogeneous and isotropic model of
FRW type a possible source of linear polarization for the CMB becomes
efficient only at the decoupling and therefore a small degree of linear
polarization seems a firmly established theoretical option  which will
be (hopefully) subjected to direct tests in the near future.
The linear polarization of the CMB is a very promising
laboratory in order to directly probe the speculated existence of a
large scale magnetic field (coherent over the horizon size at the
decoupling) which might actually rotate (through the Faraday
effect \cite{rev1,rev2,rev3}) the polarization plane of the CMB. 

Consider, for
instance, a linearly polarized electromagnetic wave of physical 
frequency $\omega$
traveling along the $\hat{x}$ direction in a cold plasma of ions and
electrons together with  a magnetic field ($ \overline{B}$)
 oriented along an arbitrary direction ( which might coincide with
$\hat{x}$ in the simplest case). 
If we let the polarization vector at the origin ($x=y=z=0$, $t=0$) 
be directed along the $\hat{y}$ axis, after the
wave has traveled a length $\Delta x$, the corresponding angular shift
($\Delta\alpha$) in the polarization plane will be :
\begin{equation}
\Delta\alpha= f_{e} \frac{e}{2m}
\left(\frac{\omega_{pl}}{\omega}\right)^2 (\overline{B}\cdot\hat{x}) \Delta x
\label{Faraday1}
\end{equation}
(conventions: $\omega_{B} = e B/m $ is the Larmor frequency;
$\omega_{pl} = \sqrt{4\pi n_{e} e^2/m}$ is the plasma frequency $n_e$
is the electron density and $f_{e}$ is the ionization fraction ;
 we use everywhere natural units $\hbar = c = k_{B}=1$).
It is worth mentioning that the previous estimate of the Faraday
rotation angle $\Delta\alpha$ holds provided $\omega\gg\omega_{B}$ and
$\omega\gg\omega_{pl}$.
From Eq. (\ref{Faraday1})
by stochastically averaging over all the possible orientations 
of $\overline{B}$  and
by assuming that the last scattering surface is infinitely thin
(i.e. that  $\Delta x f_{e} n_{e} \simeq \sigma_{T}^{-1}$ where
$\sigma_{T}$ is the Thompson cross section) we
get an expression connecting the RMS of the rotation angle to the
magnitude of $\overline{B}$ at $t\simeq t_{dec}$
\begin{equation}
\langle(\Delta\alpha)^2 \rangle^{1/2} \simeq 1.6^{0} 
\left(\frac{B(t_{dec})}{B_{c}} \right)
\left(\frac{\omega_{M}}{\omega}\right)^2,~~~B_{c} =
10^{-3}~{\rm Gauss},~~~\omega_{M} \simeq  3\times10^{10}~Hz
\label{Faraday2}
\end{equation}
(in the previous equation we implicitly assumed that the frequency of
the incident electro-magnetic radiation is centered around the maximum
of the CMB).
We can easily argue from Eq. (\ref{Faraday2}) that if $B(t_{dec}) \gaq
B_c$ the expected rotation in the polarization plane of the CMB is
non negligible.
Even if we are not interested, at this level, in a precise estimate of
$\Delta\alpha$, we point out that more refined determinations of the
expected Faraday rotation signal (for an incident frequency
$\omega_{M}\sim 30~{\rm GHz}$) were recently carried out \cite{6b,6b2}
leading to a result fairly consistent with (\ref{Faraday1}).

Then, the statement is the following. {\em If} the CMB is linearly 
polarized and {\em if} a large scale magnetic field is 
present at the decoupling epoch, {\em then} the polarization plane of the 
CMB can be rotated \cite{FR}. The predictions of different 
models  can then be confronted 
with the requirements coming from a possible detection of depolarization 
of the CMB \cite{FR}.

\renewcommand{\theequation}{9.\arabic{equation}}
\setcounter{equation}{0} 
\section{Concluding Remarks}
The large scale magnetic fields observed today in the Universe 
may or may not be primordial and there could indeed be different 
possibilities. It could be that in the past history 
of the Universe very strong magnetic fields have been created. 
These fields
could be  strong enough to affect phase transitions 
and other phenomena in the life of the Universe but, at the same time,
 too weak 
to be responsible for the origin of large scale magnetic fields.
It could also be that magnetic field were indeed strong 
enough to act as seeds of presently observed magnetic fields and, in this 
case we should be able to find evidence that this was indeed 
the case. In light of this perspective 
various ``observables'' , possibly affected by the existence 
of primordial magnetic fields, could be proposed. They include 
the stochastic GW backgrounds, the Faraday rotation of CMB and 
the baryon asymmetry of the Universe. 
From a more theoretical perspective, primordial magnetic 
fields can be connected 
to the exsistence of 
(small and large) extra-dimensions and to the 
possible dynamics of gauge couplings in the early stages of the 
evolution of the Universe. 

\Acknowledgements
The author wishes to thank Norma Sanchez and Hector 
de Vega for providing a very stimulating environment 
at the ``7th Colloque de Cosmologie'' and for interesting 
remarks. The author 
wishes also to express his gratitude 
to Joachim Tr\"umper, Daniel Boyanovsky, Rocky Kolb, Michele Simionato, 
Peter Biermann for fruitful discussions and comments.

\end{document}